\documentclass[prd,aps,floats,floatfix,superscriptaddress,preprintnumbers,
showpacs,eqsecnum,nofootinbib,twocolumn]{revtex4}
\usepackage{latexsym,array,theorem,mathrsfs,bm,float}

\usepackage{psfrag}
\usepackage{amsfonts,amsmath,amssymb,latexsym,array,afterpage,theorem,mathrsfs,bm,float,epsfig,color,graphicx,tabularx,here,multirow}

\usepackage{tikz}
\usetikzlibrary{decorations.pathmorphing,decorations.markings, patterns,shapes}
\newcommand{\nn}{\nonumber}

\newcommand{\beq}{\begin{equation}}
\newcommand{\eeq}{\end{equation}}
\newcommand{\beqy}{\begin{eqnarray}}
\newcommand{\eeqy}{\end{eqnarray}}
\newcommand{\bea}{\begin{eqnarray}}
\newcommand{\ena}{\end{eqnarray}}

\begin{document}

\title{
Particle motions and Gravitational Lensing in de Rham-Gabadadze-Tolley Massive Gravity Theory
}


\author{Sirachak {\sc Panpanich}}
\email{sirachakp-at-gmail.com}
\address{High Energy Physics Theory Group, Department of Physics, 
Faculty of Science, Chulalongkorn University, Phayathai Rd., 
Bangkok 10330, Thailand}
\author{Supakchai {\sc Ponglertsakul}}
\email{supakchai.p-at-gmail.com}
\address{High Energy Physics Theory Group, Department of Physics, 
Faculty of Science, Chulalongkorn University, Phayathai Rd., 
Bangkok 10330, Thailand}
\author{Lunchakorn {\sc Tannukij}}
\email{l_tannukij-at-hotmail.com}
\address{Department of Physics, Hanyang University, 
Seoul 133-891, South Korea}


\date{\today}

\begin{abstract}
We investigate gravitational lensing and particle motions around non-asymptotically flat black hole spacetime in non-linear, ghost-free massive gravity theory, called de Rham-Gabadadze-Tolley (dRGT) massive gravity. Deflection angle formulae are derived in terms of perihelion parameter. The deflection angle can be positive, zero or even negative with various perihelion distance. The negative angle reveals repulsive behaviour of gravity from a linear term $\gamma$ in the dRGT black hole solution. We also find an analytically approximated formula of deflection angle in two regimes: large and small $\gamma$ term regimes which are shown to be consistent with direct numerical integration. Null and timelike geodesic motions on equatorial plane are explored. Particle trajectories around the dRGT black hole are plotted and discussed in details.
\end{abstract}



\maketitle

\section{Introduction}
\label{sec:intro}

Despite general relativity (GR) has successfully described many gravitational phenomena, however it fails to explain an accelerated expansion of the Universe \cite{Riess:1998cb,Perlmutter:1998np}. This indicates that either the Universe is filled up with mysterious dark energy or some modification is needed for GR. A number of modified gravity models have been proposed to resolve an unexplained accelerated expansion. Many modified gravity theories demand a new degree of freedom e.g. a scalar field in the Horndeski theories \cite{Horndeski:1974wa,Deffayet:2011gz,Kobayashi:2011nu}, a massive vector field in generalised Proca theories \cite{Heisenberg:2014rta,DeFelice:2016yws} and a massive graviton field in de Rham-Gabadadze-Tolley (dRGT) massive gravity theory \cite{deRham:2010ik,deRham:2010kj}.

The dRGT massive gravity is a non-linear generalisation of the linear Fierz-Pauli massive gravity \cite{Fierz:1939ix}, the very first linear model which propagates massive spin-2 degrees of freedom. While this non-linear generalisation fixes the problem of van Dam-Veltman-Zakharov (vDVZ) discontinuity \cite{vanDam:1970vg,Zakharov:1970cc} -- the linear massive gravity in its massless limit does not give the same prediction as general relativity does. The dRGT massive gravity theory is also free of non-linear ghost instability called Boulware-Deser (BD) ghost which usually arises when the non-linear effect is introduced \cite{Boulware:1973my}. In dRGT massive gravity, massive graviton generates an effective cosmological constant \cite{Gumrukcuoglu:2011ew,Gumrukcuoglu:2011zh}, however, the model has a severe problem. Cosmological solutions are unstable for the Friedmann-Lema\^{\i}tre-Robertson-Walker (FLRW) background \cite{DeFelice:2012mx}.

Nevertheless, phenomenology of the dRGT massive gravity model is still interesting. Black hole (BH), black string (BS) and recently, rotating black string solutions of the dRGT model are discovered and investigated in \cite{Ghosh:2015cva,Tannukij:2017jtn,Ghosh:2019eoo,Nieuwenhuizen:2011sq}. In the presence of the massive graviton, the dRGT BH and BS solutions are modified from the conventional general relativity solution since the graviton mass contributes as a cosmological constant-like term, a linear term, and a global monopole term. 
Similar terms also arise in the minimal theory of massive gravity \cite{DeFelice:2018vza} and Bergshoeff, Hohm and Townsend (BHT) massive gravity \cite{Oliva:2009ip}. This makes dRGT solutions very interesting. Many investigations using these dRGT solutions have been done, for example, fitting rotation curves of galaxies \cite{Panpanich:2018cxo} in the presence of massive graviton instead of dark matter, stabilities and greybody factor on charged black holes and black strings \cite{Burikham:2017gdm,Ponglertsakul:2018smo,Boonserm:2017qcq,Boonserm:2019mon}.

Einstein's general relativity predicts a deflection of light traveling around massive body. A distribution of matter e.g. galaxies or supermassive black hole, can behave as a lens which distorts the light from a distant source to an observer. This effect is known as gravitational lensing phenomenon. It was first identified observationally as a double quasar system in 1979 \cite{Walsh:1979nx}. Since then, it has become one of the important research topics in gravitational physics. Gravitational lensing around black holes and compact objects have been studied in many scenarios such as Schwarzschild black hole \cite{Virbhadra:1999nm}, BHs with the presence of cosmological constant \cite{Zhao:2016ltm,Rindler:2007zz}, naked singularity and horizonless ultra-compact object \cite{Virbhadra:2002ju,Shaikh:2019itn}.

Even though, many attempts have been done to investigate the effect of gravitational lensing in different massive gravity models \cite{Jusufi:2017drg,Zhang:2018per,Nakashi:2019jjj} but none has been done for the dRGT massive gravity. Therefore, in this paper, we will study deflection angle of light from the dRGT BH solution, and analyse geodesics trajectory around the BH for both null and timelike particles.

This paper is organised as follows. In section \ref{sec:basic} we state about the basic equations and derive a photon sphere equation and deflection angle of light from a general static spherically symmetric background. In section \ref{deflectionangle}, we introduce the dRGT charged black hole. We find analytically approximated deflection angle formula for two regimes: large and small $\gamma$ term. We compare the formula with our direct numerical integration method. Gravitational lensing of supermassive black hole and galaxies modeled by the dRGT BH solution is also investigated in this section. Null and timelike geodesics are explored via analysing an effective potential in section \ref{geodesics}. Lastly, section \ref{conclusions} is devoted for our conclusions. 

In this paper, we work in the unit such that $\hbar = c=1$ unless otherwise stated.

\section{Basic Equations}
\label{sec:basic}
\allowdisplaybreaks

In this section, we shall derive generic equations for a photon sphere and deflection angle. A general static spherically symmetric spacetime is given by the following line element
\begin{align}
ds^2 = -A(r)dt^2 &+ B(r) dr^2 \nn \\
& + D(r)r^2\left(d\theta^2 + \sin^2\theta d\varphi^2\right). \label{genmetric}
\end{align}
We first study a photon trajectory in this generic spacetime by considering null geodesic equations. Without loss of generality, we focus on equatorial motion of photon, i.e. $\theta=\pi/2$. Null geodesic is then described by
\begin{align}
A \dot{t} &= E, \label{geotime}\\
2B \ddot{r} + A'\dot{t}^2 + B' \dot{r}^2 -\left(D' r^2+2D r\right)\dot{\varphi}^2 &= 0, \label{georadial}\\
D r^2 \dot{\varphi} &= L, \label{geophi}
\end{align}
where $\dot{r}\equiv dr/d\sigma$, prime denotes derivative with respect to $r$, and $\sigma$ is an affine parameter. Remark that $\theta$ component is trivially satisfied. The constants of motion are $E$ and $L$. 
In addition, the line-element of photon implies that $g_{\mu\nu}\dot{x}^{\mu}\dot{x}^{\nu}=0$. This can be shown explicitly as 
\begin{align}
-A \dot{t}^2 + B \dot{r}^2 + Dr^2\dot{\varphi}^2 &= 0. \label{geoEL}
\end{align}
Firstly, we shall consider a photon sphere. The photon sphere is defined to be a spherical region around a black hole where gravity is strong enough to force photon traveling in a circle. Therefore at the photon radius we define $r=r_{ps}, ~ \ddot{r}=\dot{r}=0$. For generic spherically symmetric spacetime, the Eqs. (\ref{georadial}) and (\ref{geoEL}) yield a photon sphere equation 
\begin{align}
\frac{A^{\prime}}{A} = \frac{D^{\prime}}{D} + \frac{2}{r}. \label{genphotosphere}
\end{align}
Note that, this equation must be evaluated at the photon sphere radius. 

\begin{figure}[h]
\vspace{-2.1cm}
 \includegraphics[width=0.45\textwidth]{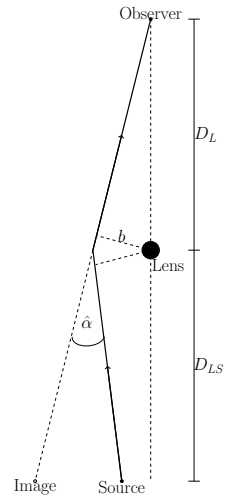}
\vspace{-3cm}
\caption{The lens diagram:  $D_{LS}$ is the distance from source to lens (deflector) whereas the distance from observer to lens is shown by $D_{L}$. The deflection angle of light is denoted by $\hat{\alpha}$. $b$ is the impact parameter.} 
\label{lensdiagram}
\end{figure}

The light bending angle or deflection angle can also be determined from the geodesic equation. To see this, one need to express $r$ as a function of $\varphi$. It is also convenient to define new radial coordinate $u\equiv1/r$. 
Therefore Eq. (\ref{geoEL}) implies the following
\begin{align}
\frac{du}{d\varphi} &= \left(\frac{D^2}{AB b^2} - \frac{Du^2}{B}\right)^{1/2}, \label{derivativeform}
\end{align}
where we have defined impact parameter as $b^2 \equiv L^2/E^2$. By integrating this equation, we obtain the following
\begin{align}
\Delta \varphi = 2 \int_{u_L}^{u_0} \left(\frac{D^2}{ABb^2} - \frac{Du^2}{B}\right)^{-1/2} du \,. \label{genAngle}
\end{align}
The factor of 2 arises from symmetry of distances between a source to the lens $(D_{LS})$ and an observer to the lens $(D_{L})$, where we assume $D_{LS} = D_L$ throughout this work. This integral is then determined from an inverse distance of the observer to the lens $(u_L = 1/D_L)$ to an inverse distance of the perihelion $(u_0 = 1/r_0)$. Note that if the observer is at infinity, then $u_L \to 0$, and the bending angle becomes very large when $u_0 \to u_{ps}$, where $u_{ps}=1/r_{ps}$, which is an inverse photon sphere radius. The lens diagram is illustrated in Fig. \ref{lensdiagram}. In this diagram, the light ray from the source is deflected by the lens (deflector) which produces a strong curvature effect on background spacetime. The deflection of light ray is measured by $\hat{\alpha}$. The impact parameter is represented by $b$. 
\

%
%
%
%
%
%
%
%
%
%
%
%
%

In addition, from Eq. (\ref{derivativeform}), the impact parameter $b$ can be defined in another way as
\begin{align}
b &= r_0\sqrt{\frac{D(r_0)}{A(r_0)}}. \label{impact}
\end{align}
It is obvious that $b \simeq r_0$ in weak gravity limit.

\section{Deflection angle of light in dRGT massive gravity}
\label{deflectionangle}

The dRGT massive gravity coupled with electromagnetic field in Gaussian units can be described by the following action \cite{deRham:2010ik,deRham:2010kj,Ghosh:2015cva}
\begin{equation}
S = \int d^4 x \sqrt{-g} \frac{M_{\rm PL}^2}{2} \left[R + m_g^2 \mathcal{U}(g,f) - F_{\mu\nu}F^{\mu\nu}\right] \,, \label{action}
\end{equation}
where $M_{\rm PL}$ is the reduced Planck mass, $R$ is the Ricci scalar, $m_g$ is the graviton mass, and $\mathcal{U}$ is self-interacting potential of the gravitons. The ghost-free self-interacting potential $U(g,f)$ is given by
\begin{align*}
&\mathcal{U} \equiv \mathcal{U}_2 + \alpha_3 \mathcal{U}_3 + \alpha_4 \mathcal{U}_4 \,, \\
&\mathcal{U}_2 \equiv  [\mathcal{K}]^2 - [\mathcal{K}^2] \,, \\
&\mathcal{U}_3 \equiv [\mathcal{K}]^3 - 3 [\mathcal{K}][\mathcal{K}^2] + 2 [\mathcal{K}^3] \,, \\
&\mathcal{U}_4 \equiv [\mathcal{K}]^4 - 6 [\mathcal{K}]^2 [\mathcal{K}^2] + 3[\mathcal{K}^2]^2 + 8 [\mathcal{K}][\mathcal{K}^3] - 6 [\mathcal{K}^4] \,,
\end{align*}
where the tensor $\mathcal{K}^{\mu}_{\nu}$ is defined as
\begin{equation}
\mathcal{K}^{\mu}_{\nu} \equiv \delta^{\mu}_{\nu} - \sqrt{g^{\mu\lambda} \partial_{\lambda} \phi^a \partial_{\nu} \phi^b f_{ab}} \,, \nonumber
\end{equation}
and $[\mathcal{K}] = \mathcal{K}^{\mu}_{\mu}$ and $(\mathcal{K}^i)^{\mu}_{\nu} = \mathcal{K}^{\mu}_{\rho_1} \mathcal{K}^{\rho_1}_{\rho_2} ... \mathcal{K}^{\rho_i}_{\nu}$.
$\phi^a$ are the St$\Ddot{\rm u}$ckelberg fields where we use the unitary gauge as $\phi^a = x^{\mu} \delta^a_{\mu}$. The field strength tensor $F_{\mu\nu}$ is defined by $\nabla_{\mu}A_{\nu}-\nabla_{\nu}A_{\mu}$. In dRGT massive gravity the physical metric is $g_{\mu\nu}$, whereas a non-dynamical (fiducial) metric $f_{\mu\nu}$ is chosen to be \cite{Vegh:2013sk,Cai:2014znn}
\begin{equation}
f_{\mu\nu} = 
  \begin{pmatrix}
    0 & 0 & 0 & 0 \\
    0 & 0 & 0 & 0 \\
    0 & 0 & C^2 & 0 \\
    0 & 0 & 0 & C^2 \sin^2 {\theta}
  \end{pmatrix} \,,
\end{equation} 
where $C$ is a positive constant.
 
Variation the action Eq. (\ref{action}) with respect to $g^{\mu\nu}$ gives the field equations
\begin{equation}
G^{\mu}_{\nu} + m^2_g X^{\mu}_{\nu} = T^{\mu (F)}_{\nu} \,, \label{EFE}
\end{equation}
where $T^{\mu (F)}_{\nu} = 2 \left(F^{\mu}_{~\rho} F_{\nu}^{~\rho} - \frac{1}{4} \delta^{\mu}_{\nu} F_{\rho\sigma}F^{\rho\sigma}\right)$ is the energy-momentum tensor of the Maxwell field. The massive graviton tensor $X^{\mu}_{\nu}$ is given by \cite{Berezhiani:2011mt,Cai:2012db,Ghosh:2015cva} 
\begin{align}
X^{\mu}_{\nu} &= \mathcal{K}^{\mu}_{\nu} - [\mathcal{K}] \delta^{\mu}_{\nu} \nonumber \\
&~~ - \alpha \left[(\mathcal{K}^2)^{\mu}_{\nu} - [\mathcal{K}]\mathcal{K}^{\mu}_{\nu} +\frac{1}{2}\delta^{\mu}_{\nu} ([\mathcal{K}]^2 - [\mathcal{K}^2])\right] \nonumber \\
&~~ + 3 \beta \left[(\mathcal{K}^3)^{\mu}_{\nu} - [\mathcal{K}](\mathcal{K}^2)^{\mu}_{\nu} +\frac{1}{2}\mathcal{K}^{\mu}_{\nu} ([\mathcal{K}]^2 - [\mathcal{K}^2])\right. \nonumber \\
&~~ \left. - \frac{1}{6} \delta^{\mu}_{\nu} ([\mathcal{K}]^3 - 3 [\mathcal{K}][\mathcal{K}^2] + 2[\mathcal{K}^3]) \right] \,, \nonumber
\end{align}

where $\alpha_3 = \frac{\alpha - 1}{3}$, $\alpha_4 = \frac{\beta}{4} + \frac{1 - \alpha}{12}$.

By using the ansatz $A_{\mu}=(A(r),0,0,0)$, the static spherically symmetric black hole solution of Eq. (\ref{EFE}) is given by \cite{Ghosh:2015cva}
\begin{align}
ds^{2} &= -f(r)dt^2 + \frac{1}{f(r)}dr^2 + r^2d\theta^2 + r^2\sin^2\theta d\varphi^2, 
\end{align}
where 
\begin{align}
f(r) &= 1 - \frac{2GM}{r} + \frac{Q^2}{r^2} - \frac{\Lambda}{3}r^2 + \gamma r + \zeta. \label{drgtmetric} 
\end{align}
The mass and electric charge of the black hole are denoted by $M$ and $Q$, respectively. The constant $\Lambda,\gamma$ and $\zeta$ effectively emerge from the graviton mass which is the unique character of the dRGT black hole solution, i.e. 
\bea
\Lambda &=& -3m^2_g(1+\alpha+\beta), \label{lamb} \\
\gamma &=& -Cm^2_g(1+2\alpha+3\beta), \label{gam} \\
\zeta &=& C^2m^2_g(\alpha+3\beta) .
\ena
Due to the presence of $r^2$ and linear $r$, an asymptotic structure changes correspondingly with the signs of $\Lambda$ and $\gamma$. However, we shall only consider in the case where both $\Lambda$ and $\gamma$ are positive in this work. 

In comparison with the general metric Eq. $(\ref{genmetric})$, hence we obtain
\begin{align}
A = B^{-1} = f,~~~~~~~ D \equiv 1.
\end{align}
Substituting the metric function Eq. (\ref{drgtmetric}) into the photon sphere equation, Eq. (\ref{genphotosphere}), we find
\begin{align}
\gamma r_{ps}^3 + 2 (1 + \zeta) r_{ps}^2 - 6GMr_{ps} + 4Q^2 &= 0. \label{photoneq1}
\end{align}
This equation correctly yields $r_{ps}=3GM$ in the Schwarzschild metric limit (i.e. $Q = 0$, $\Lambda = \gamma = \zeta = 0$). It can be seen that the photon sphere equation does not depend on the cosmological constant. The radius of photon sphere is then determined from the roots of this equation. 
Since this is cubic equation, it is possible to obtain three real roots. We shall choose only the one where $r_h<r_{ps}<r_c$ where $r_h$ and $r_c$ refer to the event horizon and the cosmological horizon of a black hole. In addition, we analyse root structure of Eq. (\ref{drgtmetric}) and display some parameter spaces in Appendix \ref{chargedBH}.


Now Eq. (\ref{genAngle}) can be expressed in terms of the dRGT metric Eq. (\ref{drgtmetric}) 

\begin{widetext}
\bea
{\displaystyle 
\Delta \varphi = 2 \int_{u_L}^{u_0} \frac{du}{\sqrt{-Q^2 u^4 + 2GM u^3 - (1 + \zeta) u^2 - \gamma u + \frac{1}{b^2} + \frac{\Lambda}{3}}} \,. \label{drgtangle}
}
\ena
We assume that the distances between the source to the lens and the observer to the lens are equal, i.e. $D_L = D_{LS}$, thus $u_L = u_{LS}$. Since the source and the observer in practice are not at infinity, the limits of integration must be from $u_L$ to $u_0$, which are an inverse distance of the observer to the lens ($u_L = 1/D_L$) and an inverse distance of perihelion ($u_0 = 1/r_0$), respectively. Using the fact that at perihelion $du/d\varphi = 0$, then we can rewrite Eq. (\ref{drgtangle}) as 
\bea
\Delta \varphi = 2 \int_{u_L}^{u_0} \frac{du}{\sqrt{Q^2 (u_0^4 - u^4) + 2GM (u^3 - u_0^3) + (1 + \zeta) (u_0^2 - u^2) + \gamma (u_0 - u)}} \,. \label{drgtangle2}
\ena
\end{widetext}
The cosmological constant and the impact parameter are absorbed in terms of the perihelion distance, $u_0$. In the absence of massive graviton, i.e. $\gamma=\zeta=0$, the $u^2$ term becomes dominant at large distance (hence small $u$). However, when $\gamma$ and $\zeta$ are non-vanishing, there is a possibility that the $\gamma$ term will dominate at some distance. Hence, we shall derive analytic formula of the deflection angle in two possible regimes:  a small $\gamma$ term regime and a large $\gamma$ term regime. To clarify the terminology, here a small $\gamma$ regime refers to the case where the linear term in $u$ is less than the $u^2$ term. A large $\gamma$ regime, on the other hand, represents the case where $u^2$ term is comparatively smaller than the linear $\gamma$ term.

It is worth mentioning that the above integral is a total angle that light passes by from the source to the observer in non-asymptotically flat spacetime, thus the bending angle or deflection angle is actually given by 
\bea
\hat \alpha = \Delta \varphi + \varphi_0 - \varphi_s \,, \label{drgtangle1}
\ena
where 
\bea
\varphi_0 &=& \arccos \left[\sqrt{1 - \frac{r_0^2}{D_L^2} \frac{f(D_L)}{f(r_0)}}\right] \,, \\
\varphi_s &=& \arccos \left[- \sqrt{1 - \frac{r_0^2}{D_{LS}^2} \frac{f(D_{LS})}{f(r_0)}}\right] \,,
\ena
which can be obtained from geometrical analysis (see Refs. \cite{Ishihara:2016vdc,Zhao:2016ltm,Rindler:2007zz}). In an asymptotically flat spacetime, the term $\varphi_0 - \varphi_s = - \pi$ as in conventional GR. 

\subsection{Large $\gamma$ term regime}

If the $\gamma u$ term dominates, while other terms are smaller than the $u^2$ term  (we also assume $\zeta \ll 1$), we can rewrite Eq. (\ref{drgtangle2}) as
\begin{widetext}
\bea
\Delta \varphi = 2 \int_{u_L}^{u_0} \frac{du}{\sqrt{\gamma u_0 (1 - \tilde u)}} \left\{ 1 + \frac{(1 + \zeta)u_0}{\gamma} (1 + \tilde u) - \frac{2GMu_0^2}{\gamma} (1 + \tilde u + \tilde u^2) + \frac{Q^2 u_0^3}{\gamma} (1 + \tilde u + \tilde u^2 + \tilde u^3)\right\}^{-1/2} \,, \nn \\
\ena
where $\tilde u \equiv u/ u_0$. Thus, the analytic solution can be approximated by using the binomial expansion as 
\bea
\hat \alpha &=& 2 \sqrt{\frac{1}{\gamma r_0}} \sqrt{1- \frac{r_0}{D_L}} \left\{2 - \frac{(1+\zeta)}{3\gamma r_0}\left(5+\frac{r_0}{D_L}\right)
+ \frac{2 GM}{5\gamma r_0^2} \left(11 + 3\frac{r_0}{D_L} + \left(\frac{r_0}{D_L}\right)^2 \right) \right. \nn \\
& & \left. - \frac{Q^2}{35 \gamma r_0^3} \left(93 + 29\frac{r_0}{D_L} +13 \left(\frac{r_0}{D_L}\right)^2 + 5\left(\frac{r_0}{D_L}\right)^3\right)\right\}  - \pi + 2 \arccos\left[\sqrt{1 - \frac{r_0^2}{D_L^2} \frac{f(D_L)}{f(r_0)}}\right] \,. ~~~ \label{analytic2}
\ena
\end{widetext}
Notice that, we express the above formula in term of $r_0$ and $D_L$. This expression helps us to visualize
deflection angle formula better.

\begin{figure}[t]
\centering
 \includegraphics[width=0.45\textwidth]{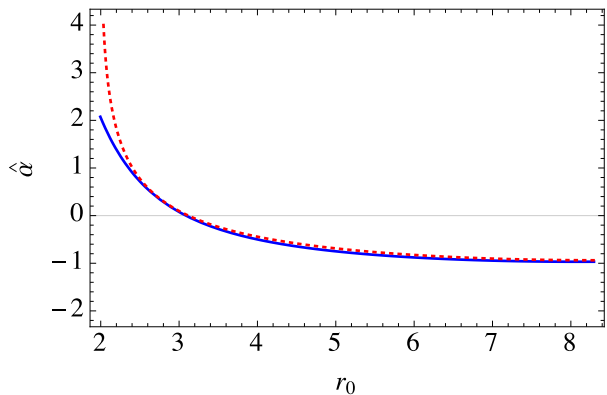}
\caption{The comparison between analytic solution and numerical solution of the deflection angle of light where we set $G = M=1, Q=0.1,\Lambda=0.1$, and $\gamma=0.5$. The blue line represents the analytic solution Eq. (\ref{analytic2}), whereas the red-dotted line represents the numerical solution of Eq. (\ref{drgtangle1}).}
\label{strongangle}
\end{figure}

The numerical solution for deflection angle Eq. $(\ref{drgtangle1})$ and the analytic solution Eq. $(\ref{analytic2})$ as a function of $r_0$ are compared in Fig. \ref{strongangle}. The perihelion $r_0$ ranges from the photon sphere radius $r_{ps}$ to the lens-observer distance $D_L$ namely, $r_h<r_{ps} < r_0 < D_L < r_c$. We set $r_0 ({\rm max}) = 2D_L/3$, $D_L = 3 r_c /4$, and $\zeta = 0$. It is clear that they agree with great accuracy. The deviation between these two methods seems to occur only as $r_0\to r_{ps}$ because it does not satisfy the large $\gamma$ term approximation.

\begin{figure}[t]
\centering
\includegraphics[width=0.45\textwidth]{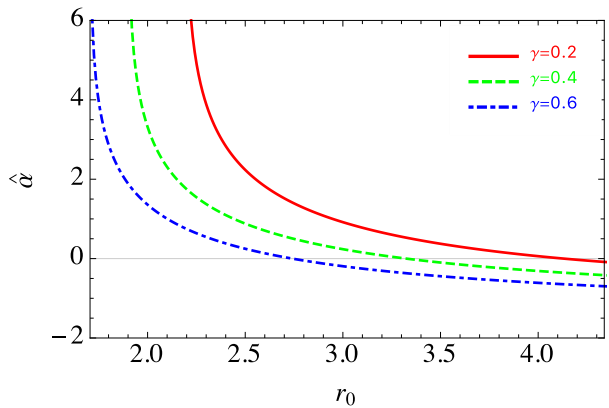}
\caption{Example plots of deflection angle as a function of $r_0$, where we set $G = M=1,Q=0.6,\Lambda=0.1$ and vary $\gamma$ parameter.}
\label{varygamma}
\end{figure}

We explore the deflection angle for various values of $\gamma$ numerically in Fig.~\ref{varygamma}. We basically plot Eq. $(\ref{drgtangle1})$ as a function of $r_0$, where we set $\zeta = 0$, and $r_0 ({\rm max}) = 2D_L/3$, $D_L = 3 r_c /4$. As can be seen from these plots, the deflection angles diverge as $r_0\to r_{ps}$, and then decrease as $r_0$ increases.

From these figures, the light bending angle can be either positive, zero and negative. The positive angle shows standard attractive effect of Einstein's gravity. Surprisingly, we observe that deflection angle becomes negative for certain value of $r_0$. This can be interpreted as repulsive effect of gravity which is one of the unique characters of the dRGT massive gravity.
It should be remarked that, negative deflection angle is also found in the BHT massive gravity \cite{Nakashi:2019jjj} and in modified gravity theories with exotic matter and energy \cite{Kitamura:2012zy,Izumi:2013tya,Kitamura:2013tya,Nakajima:2014nba,Shaikh:2017zfl}.

\subsection{Small $\gamma$ term regime}

With $u^2$ term dominates, Eq. (\ref{drgtangle2}) can be rewritten as
\begin{widetext}
\bea
\Delta \varphi = 2 \int_{u_L}^{u_0} \frac{du}{\sqrt{u_0^2 - u^2}} \left\{1 + Q^2 (u_0^2 + u^2) - 2G M \frac{(u^2 + u u_0 + u_0^2)}{u + u_0} + \frac{\gamma}{u_0 + u} + \zeta\right\}^{-1/2} \,.
\ena
We then apply the binomial expansion and integrate the above equation directly. Hence, the deflection angle in small $\gamma$ regime is given by 
\bea
\hat \alpha &=& 2 \arccos\left(\frac{r_0}{D_L}\right) + \frac{2G M}{r_0} \left[\sin\left(\arccos\left(\frac{r_0}{D_L}\right)\right) + \tan\left(\frac{1}{2}\arccos\left(\frac{r_0}{D_L}\right)\right)\right] \nn \\
& & - \frac{Q^2}{r_0^2} \left[\frac{3}{2}\arccos\left(\frac{r_0}{D_L}\right) + \frac{1}{4}\sin\left(2\arccos\left(\frac{r_0}{D_L}\right)\right)\right] - \gamma r_0 \tan \left(\frac{1}{2}\arccos\left(\frac{r_0}{D_L}\right)\right) - \zeta \arccos\left(\frac{r_0}{D_L}\right) \nn \\
& & - \pi + 2 \arccos\left[\sqrt{1 - \frac{r_0^2}{D_L^2} \frac{f(D_L)}{f(r_0)}}\right] \,. \label{analytic1}
\ena
\end{widetext}
In the limit where graviton becomes massless $m_g\to0$, i.e. $\Lambda = \gamma = \zeta = 0$, and assumes that BH possesses no electric charge, Eq. (\ref{analytic1}) reduces to 
\bea
\hat \alpha = \frac{4GM}{r_0} \,,
\ena
with $D_L \rightarrow \infty$ (the observer is at significantly large distant from the lens). We can use the fact that $b \simeq r_0$ in weak gravity limit or small deflection angle of light. This is the conventional deflection angle in GR. 

\begin{figure}
\centering
\includegraphics[width=0.45\textwidth]{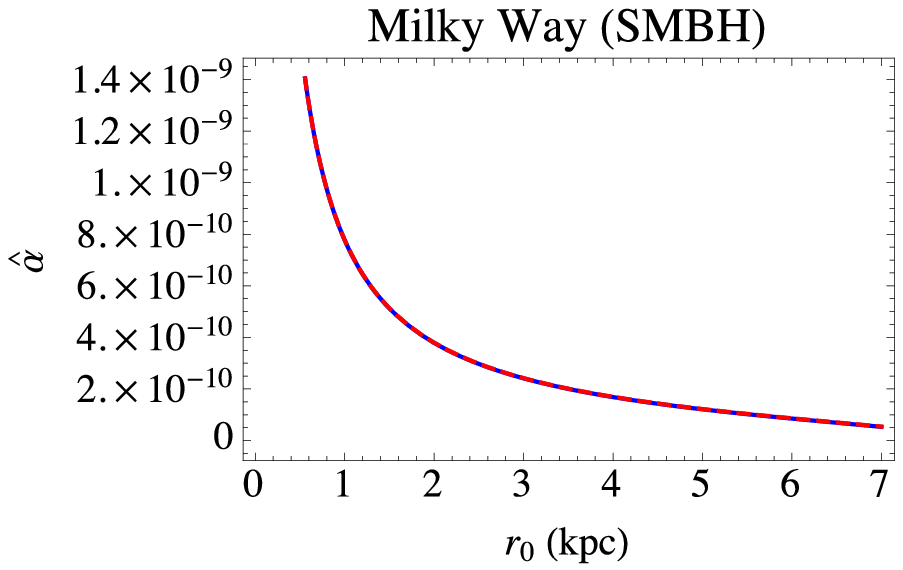} \\
\includegraphics[width=0.45\textwidth]{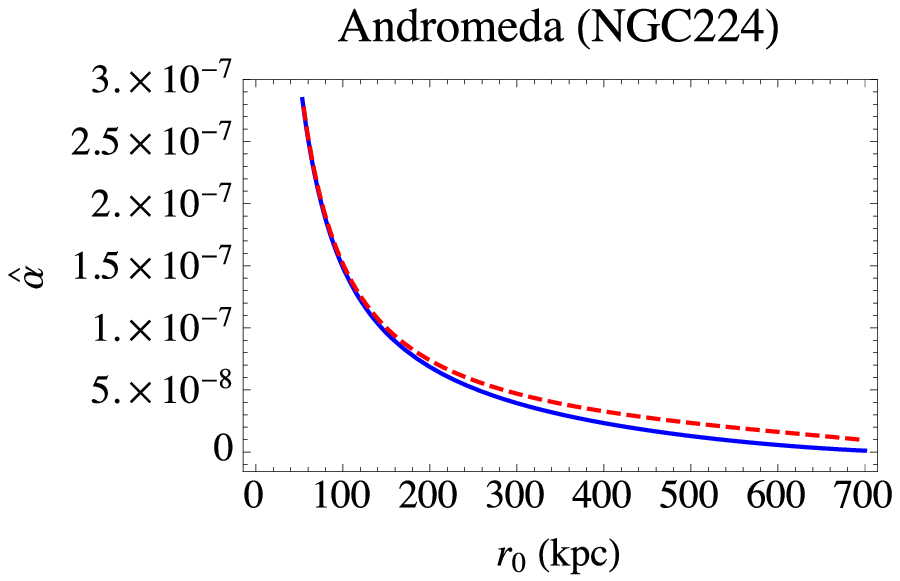} \\
\includegraphics[width=0.45\textwidth]{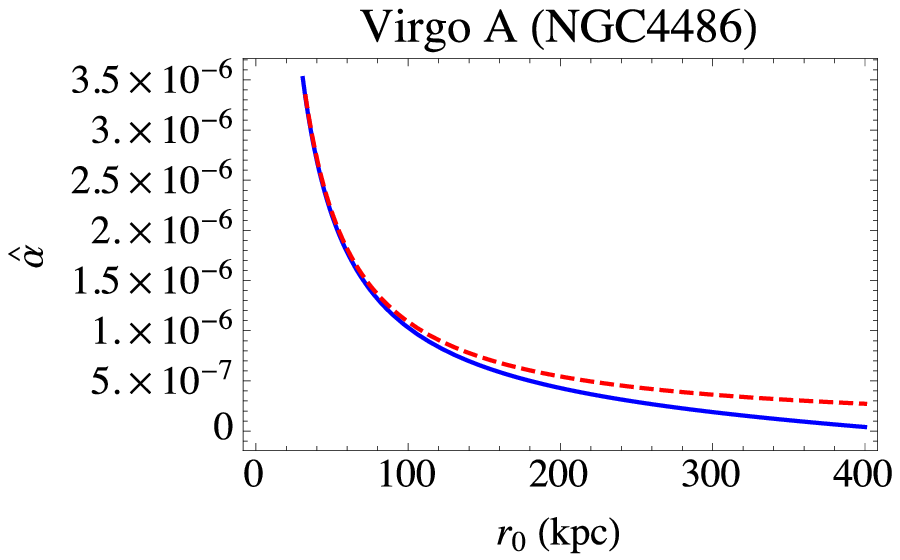}
\caption{Example plots of the deflection angle as a function of $r_0$ according to Eq. (\ref{analytic1}). Blue line is the deflection angle in the presence of dRGT massive graviton, whereas red-dashed line represents the deflection angle without the massive graviton (GR case).}
\label{weakregime}
\end{figure}

The deflection angles of light in a small $\gamma$ term regime are shown in Fig. \ref{weakregime}. In this figure, we illustrate the difference between the deflection angles of light in dRGT massive gravity and those of GR. We use observational data as displayed in Table \ref{data}. The spacetime background quantities are chosen to be $Q = 0$, $\Lambda = 1.11 \times 10^{-52} {\rm m^{-2}}$ (dark energy observations \cite{Ade:2015xua}), $\zeta = 0$, and $\gamma = 10^{-28} {\rm m^{-1}}$. This choice of parameters is motivated from the previous study where rotation curves of galaxies is explained by the presence of massive graviton instead of cold dark matter \cite{Panpanich:2018cxo}. Therefore in these plots, only visible mass in the galaxies will be taken into account which contributes approximately  $10\%$ of the total mass of the galaxies. Remark that, we are now considering the dimensionful deflection angle formula, the speed of light $c$ is no longer unity.

In Fig. \ref{weakregime}, supermassive black hole at the centre of the Milky Way galaxy (Top), Andromeda galaxy (Middle) and Virgo A galaxy (Bottom) are modeled by spherically symmetric solution in dRGT massive gravity \cite{Ghosh:2015cva}. The deflection angle by the SMBH reveals that there is no difference between dRGT massive gravity and GR. This is not unexpected since the $\gamma$ term does not dominate at this scale ($\sim 8$ kpc). This can be understood more explicitly by considering Eq. (\ref{analytic1}). It is obvious that the BH mass term yields attractive force, while the other terms provide repulsive force because of the negative signs. Since $Q=\zeta=0$, therefore, the distinct feature of dRGT or repulsive behaviour arises from the linear $\gamma$ term merely.

We assume that a galaxy is a point mass which can be described by the dRGT spherically symmetric solution. The difference between the dRGT massive gravity and GR arises when the perihelion $r_0$ is greater than $200$ kpc and $100$ kpc for Andromeda and Virgo A galaxies respectively. This is because the $\gamma$ term is relatively large at such scale, thus deflection angle decreases faster than the GR case. Our further investigation suggests that deflection angle becomes negative at the larger distance i.e. (approximately) $700$ kpc and $400$ kpc for the Andromeda galaxy and the Virgo A galaxy, respectively.

\begin{table}[t]
\begin{center}
  \begin{tabular}{|c||c|c|c|}
\hline 
& & & \\[-.5em]
Lens&Mass ($M_{\odot}$)&Distance $D_L$ & Refs. \\
& & (Mpc) &
\\[.5em]
\hline
& & & \\[-.5em]
Milky Way (SMBH) & $4.1 \times 10^6 $ & 0.008 & \cite{Ghez:2008ms}
\\[.5em]
\hline
& & & \\[-.5em]
Andromeda (NGC224)& $0.8 \times 10^{12}$ & $0.780$ & \cite{Kafle:2018amm}
\\[.5em]
\hline
& & & \\[-.5em]
Virgo A (NGC4486)& $5.7 \times 10^{12}$ & $16.4$ & \cite{Murphy:2011yz,Bird:2010rd}
\\[.5em]
\hline

 \end{tabular}
    \caption{Observational data of the supermassive black hole (SMBH) at the centre of the Milky Way, the Andromeda galaxy (NGC224), and the Virgo A galaxy (NGC4486), where the distance $D_L$ is a distance from the Earth to the lens.}
\label{data}
\end{center}
\end{table}

\begin{table}[t]
\begin{center}
  \begin{tabular}{|c||c|c|c|}
\hline 
& & & \\[-.5em]
Lens& $\left|-\frac{\Lambda r_0^2}{3} \right|$ & $\gamma r_0$ & $r_0$  \\
& & & (kpc) 
 \\[.5em]
\hline
& & & \\[-.5em]
Milky Way (SMBH) & $1.73 \times 10^{-12} $ & $2.16 \times 10^{-8}$ & $7$
\\[.5em]
\hline
& & & \\[-.5em]
Andromeda (NGC224)& $1.73 \times 10^{-8} $ & $2.16 \times 10^{-6}$ & $700$
\\[.5em]
\hline
& & & \\[-.5em]
Virgo A (NGC4486)& $5.65 \times 10^{-9} $ & $1.23 \times 10^{-6}$ & $400$
\\[.5em]
\hline

 \end{tabular}
    \caption{The absolute values of the $-\Lambda r^2 / 3$ term and the values of linear $\gamma r$ term of each lens at the large distances, where $\Lambda = 1.11 \times 10^{-52} {\rm m^{-2}}$ and $\gamma = 10^{-28} {\rm m^{-1}}$.}
\label{weakparameters}
\end{center}
\end{table}

\begin{figure}[t]
\centering
 \includegraphics[width=0.45\textwidth]{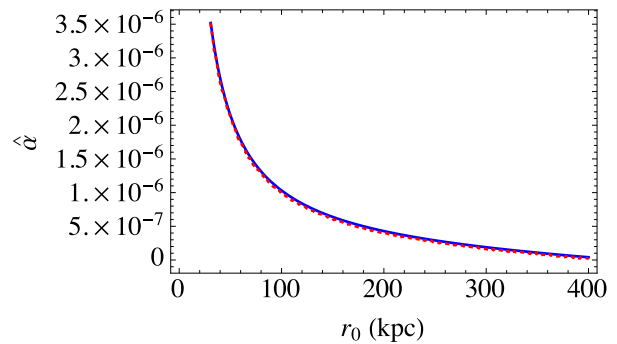}
\caption{The comparison between analytic solution and numerical solution of the deflection angle of light where we set parameters the same as the plot of Virgo A galaxy. The blue line represents the analytic solution of Eq. (\ref{analytic1}), whereas the red-dotted line represents the numerical solution of Eq. (\ref{drgtangle1}).}
\label{smallnumer}
\end{figure}


In Table \ref{weakparameters}, we show the validity of our formula in the small $\gamma$ regime. At the large distance, the $\gamma$ term is relatively larger than the cosmological constant term. Nevertheless, these two terms are still smaller than the unity. This confirms the validity of Eq. (\ref{analytic1}). Finally, we compare the deflection angle between the analytic solution of Eq. (\ref{analytic1}) and the numerical solution of Eq. (\ref{drgtangle1}) for the Virgo A galaxy in Fig. \ref{smallnumer}.  The result shows that they are consistent with high accuracy. 

It is worth mentioning that from the results on each cases, it is possible to estimate a radius at which massive gravity contributes effectively, also known as Vainshtein radius \cite{Vainshtein:1972sx}. In particular, the Vainshtein radius is a radius where the mass term and the $\Lambda,\gamma$ terms in Eq. (\ref{drgtmetric}) are comparable (we have set $\zeta=0$). In a dimensionful notation, the Vainshtein radius of the SMBH in the Milky Way galaxy can be estimated to be around an order of $10$ kpc while for the Andromeda galaxy and the Virgo A galaxy, the corresponding Vainshtein radii can be determined to be around an order of $100$ kpc. These Vainshtein radii are in agreement with the results in Fig. \ref{weakregime} where the deflection angles around the SMBH at a radius smaller than its Vainshtein radius do not differ much from GR while the deflection angles around the galaxies at the radius much larger than their Vainshtein radii significantly differ from the predictions in GR.

\section{Null and timelike Geodesics}
\label{geodesics}

In the previous sections, we have investigated the deflection angle of photon around the dRGT spherically symmetric charged black hole. In this section, rather than considering only a null geodesic, we develop a generic formula of geodesics on the background of the dRGT spherically symmetric charged black hole which includes null and timelike trajectories (for a massive particle). Without loss of generality, we consider null and timelike trajectories on a plane $\theta=\pi/2$ given by the following equation,
\begin{align}
e = -f \dot{t}^2 + \frac{\dot{r}^2}{f} + r^2 \dot{\varphi}^2, \label{eequation}
\end{align}
where $e=0,-1$ corresponds to null and timelike trajectories, respectively. Remark that, the above equation is a generalisation of Eq. (\ref{geoEL}), where we have used $A = B^{-1} = f$ (dRGT BH solution) and $D = 1$. The corresponding geodesic equations can be immediately read out from the Eqs. (\ref{geotime})-(\ref{geophi}) as follows,
\begin{align}
f \dot{t} &= E, \label{geotimea}\\
2f^{-1} \ddot{r} + f'\dot{t}^2 - f' f^{-2}\dot{r}^2 -2 r\dot{\varphi}^2 &= 0,\label{georadiala}\\
 r^2 \dot{\varphi} &= L.\label{geophia}
\end{align}
The $r-$geodesic in Eq. (\ref{georadiala}) can be simplified using Eqs. (\ref{eequation}), (\ref{geotimea}) and (\ref{geophia}) as follows,
\begin{align}
\ddot{r}-\frac{ef'}{2}+\frac{f'L^2}{2r^2}-\frac{fL^2}{r^3}=0. \label{georadial2}
\end{align}
Note that, the solution to this equation is a function of the affine parameter $\sigma$. To determine the shape of the trajectory, it is more convenient to express $r$ as a function of $\varphi$. To this end, we use the relation, $\frac{d}{d\sigma}=\frac{L}{r^2}\frac{d}{d\varphi}$ on the $r-$geodesic in Eq. (\ref{georadial2}). Eventually, the $r-$geodesic in terms of $\varphi$ can be expressed as
\begin{align}
\frac{d^2r(\varphi)}{d\varphi^2}-\frac{ef'}{2L^2}r^4-\frac{2ef}{L^2}r^3 - \frac{2E^2}{L^2}r^3 + \frac{f'r^2}{2} + fr=0.
\end{align}
In order to solve this second-order differential equation, we need two initial conditions, namely, $r(0)$ and $\frac{dr(0)}{d\varphi}$. When $r(0)$ is specified, $\frac{dr(0)}{d\varphi}$ can be found using Eq. (\ref{eequation}). However, $r(0)$ cannot be chosen arbitrarily. To determine $r(0)$, it is useful to firstly gain some knowledge about the behaviour of null and timelike trajectories through an effective potential analysis. 

\begin{figure*}[]
\includegraphics[scale=0.9]{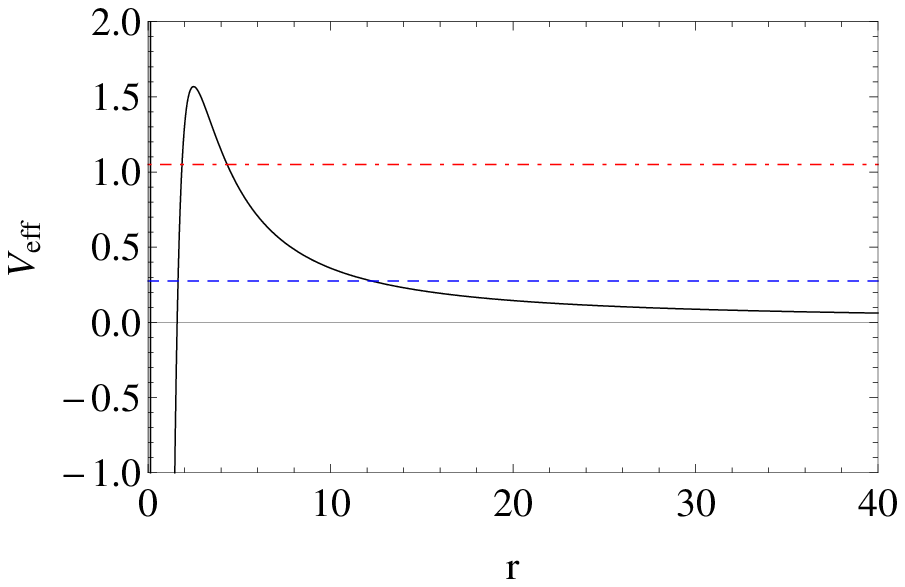}
\\
\includegraphics[scale=0.55]{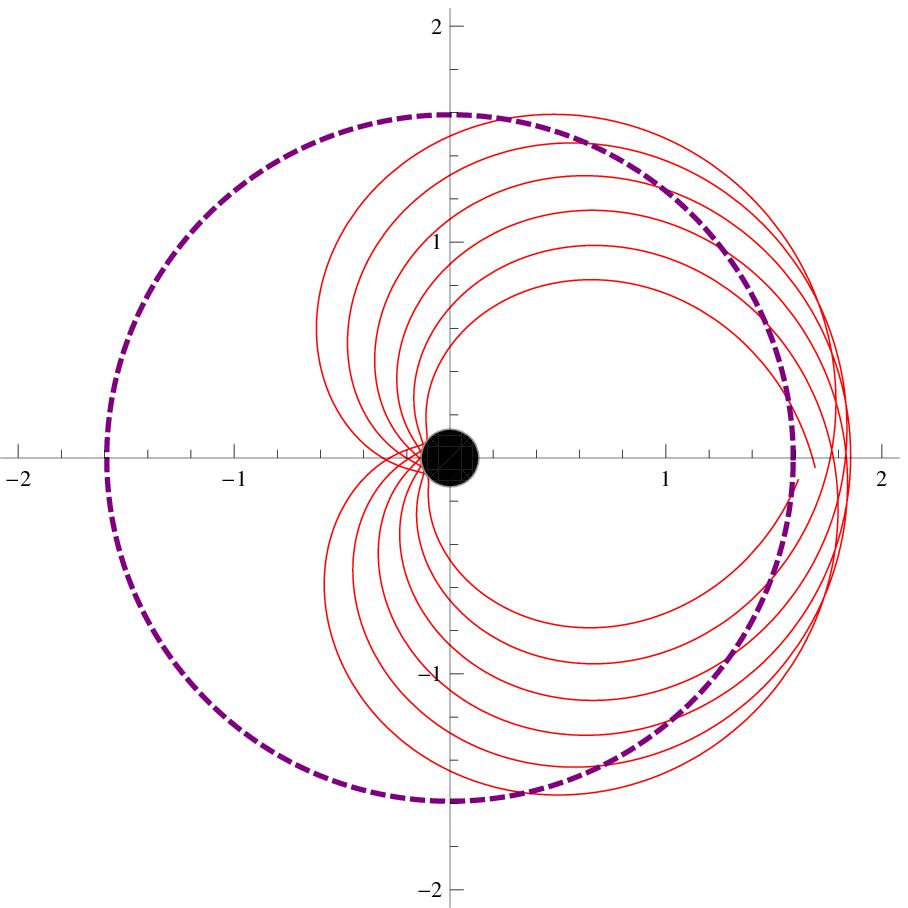}\qquad
\includegraphics[scale=0.55]{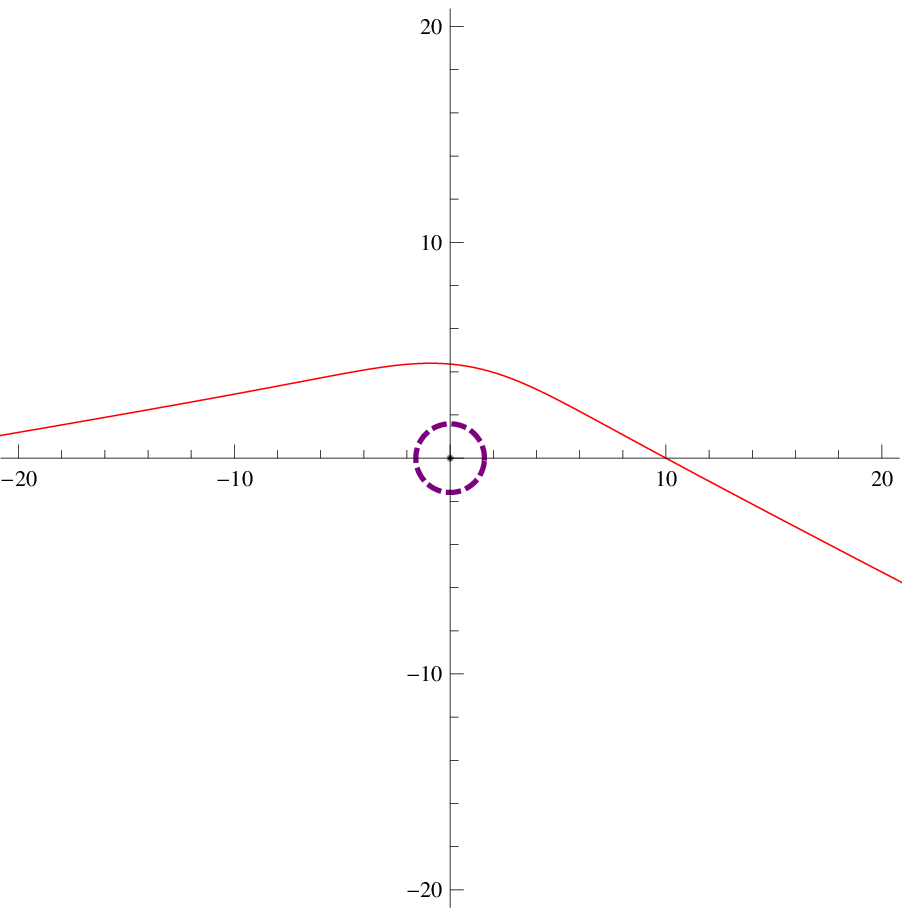}\qquad
\includegraphics[scale=0.55]{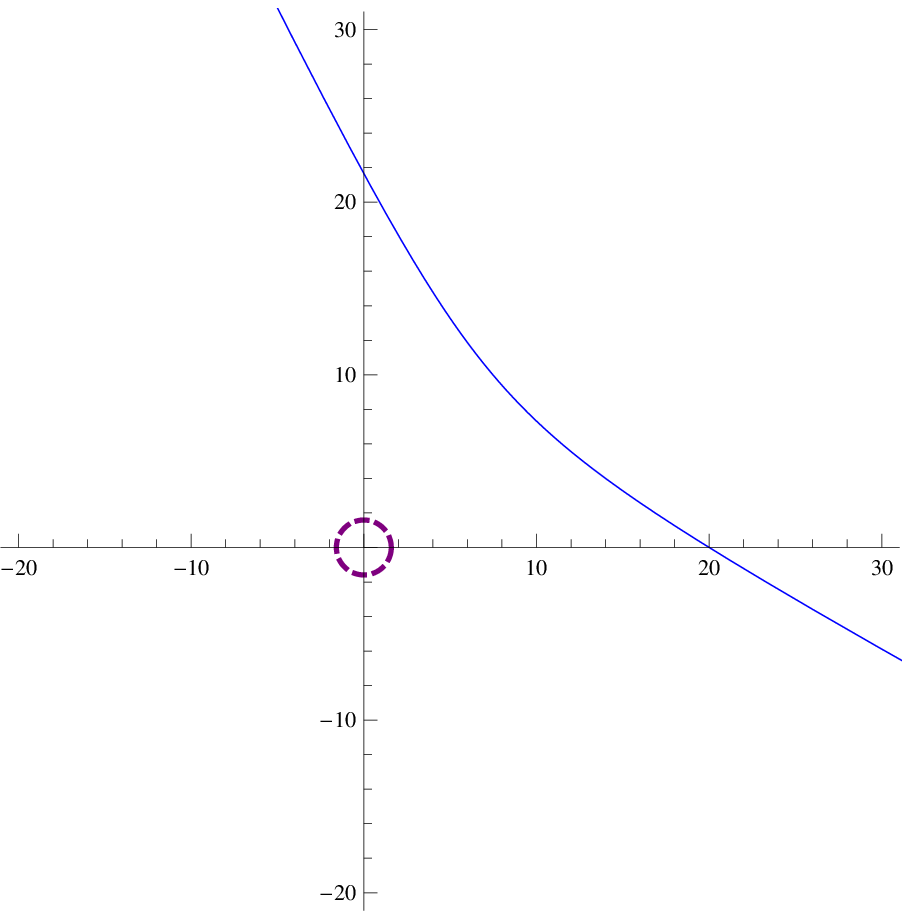}
\caption{Examples of possible null trajectories corresponding to a particular profile of $V_{eff}$. Top: the black line corresponds to the effective potential with fixed $G = M=1,Q=0.5,\Lambda=10^{-5},\gamma=0.1,L^2=40$. $E_{tot}$ is shown by two horizontal lines while dashed blue line corresponds to $E^2=0.55$ and dotted-dashed red line is $E^2=2.1$.  Bottom-left: a bounded orbit and its precession corresponding to $E^2=2.1$. Bottom-centre: an attractive fly-by trajectory with $E^2=2.1$. Bottom-right: a repulsive fly-by trajectory with $E^2=0.55$. The dashed purple circle indicates the outer event horizon of the black hole while the black circle indicates the inner event horizon.}
\label{fig:Nullfig}
\end{figure*}

From Eq. (\ref{eequation}), we can rewrite the equation into the following form,
\begin{align}
\dot{r}^2-ef-f^2 \dot{t}^2+r^2 f \dot{\varphi}^2=0. \label{conserveeq}
\end{align}
Using the explicit form of $f$ in Eq. (\ref{drgtmetric}),
we can rewrite Eq. (\ref{conserveeq}) into an ``energy conservation'' form as
\begin{align}
\frac{\dot{r}^2}{2}+V_{eff} = E_{tot}, \label{conserveeq2}
\end{align}
where
\begin{widetext}
\begin{align}
V_{eff} &\equiv \frac{e}{2}\left(\frac{2GM}{r}-\frac{Q^2}{r^2}+\frac{\Lambda r^2}{3}-\gamma r\right) + \frac{L^2}{2r^2}\left(1-\frac{2GM}{r}+\frac{Q^2}{r^2}+\gamma r +\zeta\right), \label{Veff}
\\
E_{tot} &\equiv \frac{E^2}{2}+\frac{e\left(1+\zeta\right)}{2}+\frac{\Lambda L^2}{6}. \label{Etot}
\end{align}
\end{widetext}
From Eq. (\ref{conserveeq2}), it is obvious that the allowed initial condition for $r$ must be chosen such that $V_{eff}(r)\leq E_{tot}$. Moreover, considering Eq. (\ref{Veff}), we can see how the $\gamma$ term contributes to each kind of the trajectories. For null geodesic, i.e. $e=0$, the $\gamma$ term contributes to the effective potential with $1/r$ factor, suggesting that the corresponding force is repulsive. On the other hand, a massive particle, where $e=-1$, experiences not only the repulsive $1/r$ part but is also affected by the linear term, $+\gamma r$, which contributes to the effective potential as an attractive force. Furthermore, for both photon and massive particle, the repulsive behaviour gets stronger as the angular momentum $L$ increases. This repulsive force is clearly evident in Fig. \ref{fig:Nullfig} for the photon case where the corresponding deflection angle is negative. 

There are various types of trajectories for both null and timelike cases. Each types of trajectory is determined by the initial value of $r(0)$ which associates to the location of the maximum of $V_{eff}$. Let's suppose that maximum of $V_{eff}$ is at $r_{max}$. For null case, trajectories of photon can be classified into three types. The first type is a bounded orbit which happens at $r(0)<r_{max}$. This is displayed in the lower left panel of Fig. \ref{fig:Nullfig}. In this case, the photon travels through the outer horizon into the inner horizon. Once it reaches the inner horizon, the photon is reflected by a repulsive singularity of the charged black hole. Therefore the photon crosses the inner horizon and outer event horizon in the opposite direction and enters a new copy of the static spherically symmetric spacetime. By following the photon trajectory further, the photon will repeatedly enters and comes out of the charged black hole in an infinite sequence. This feature can be more understood by considering the Carter-Penrose diagram of the Reissner-Nordstr\"{o}m solution \cite{Carroll:2004st}. In addition, this type of orbit is also found in several studies on particle motion around charged black hole spacetime, e.g. \cite{Hackmann:2008tu,Grunau:2010gd}

The second type is an attractive fly-by trajectory. This occurs when $r(0)>r_{max}$ and $E_{tot}$ must be large enough. An example of this trajectory can be found in the lower centre panel of Fig. \ref{fig:Nullfig} which corresponds to the dotted-dashed red line in the $V_{eff}$ plot. In this case, the black hole behaves as a convex lens to the photon, or the deflection angle is positive. The third type is a repulsive fly-by trajectory. This happens when $r(0)>r_{max}$ and $E_{tot}$ is relatively small. In this case, the black hole behaves as a concave lens and the photon is repelled away as shown in the lower right panel of Fig. \ref{fig:Nullfig}.


\begin{figure*}
\includegraphics[scale=0.9]{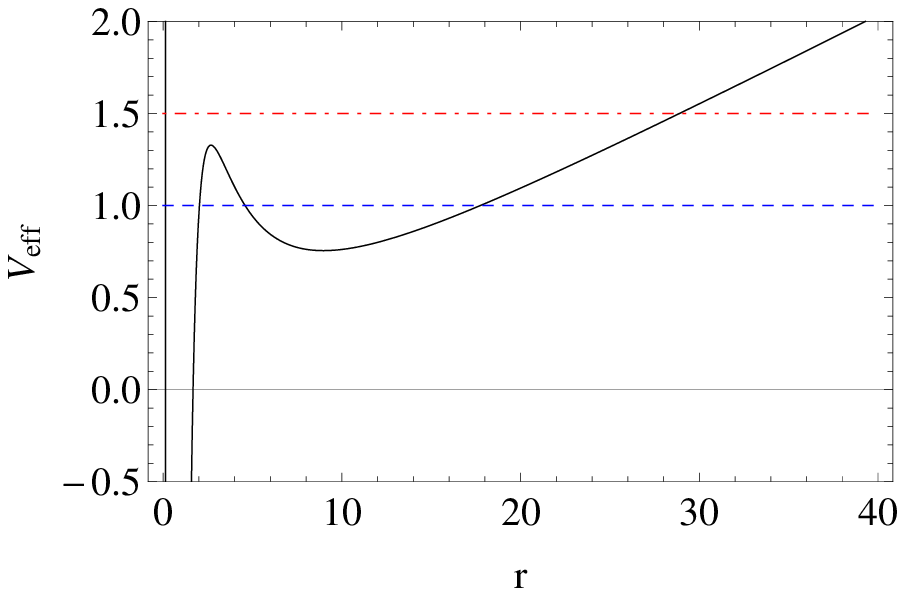}
\\
\includegraphics[scale=0.55]{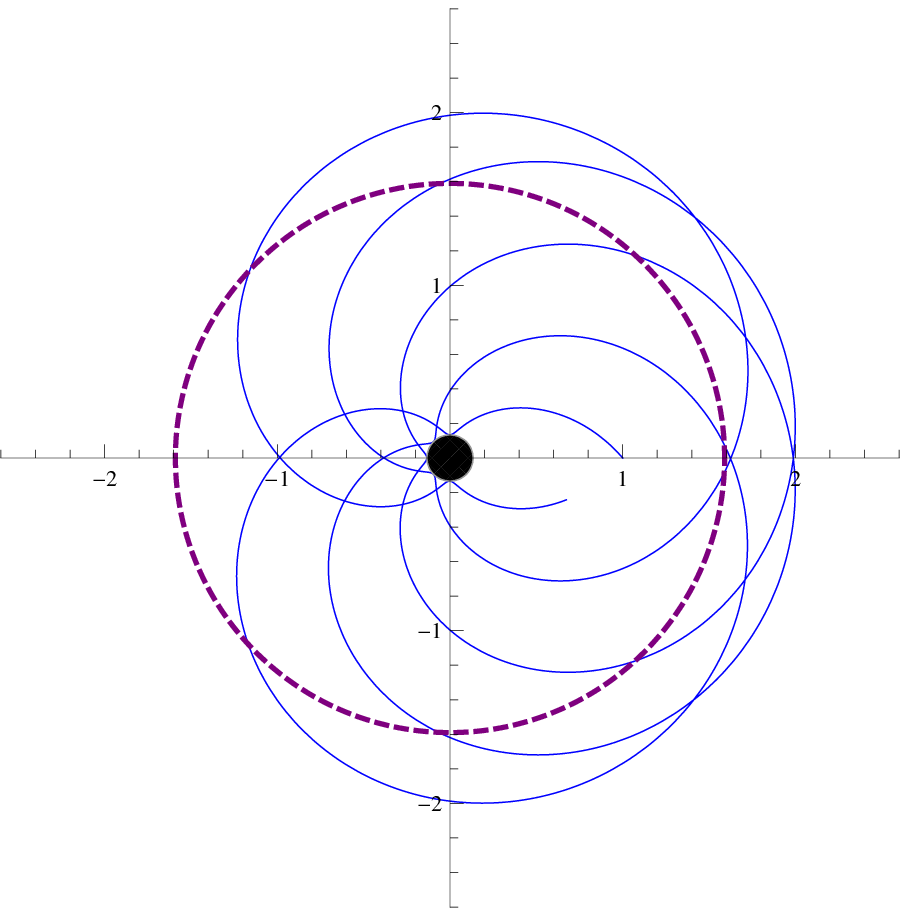}\qquad
\includegraphics[scale=0.55]{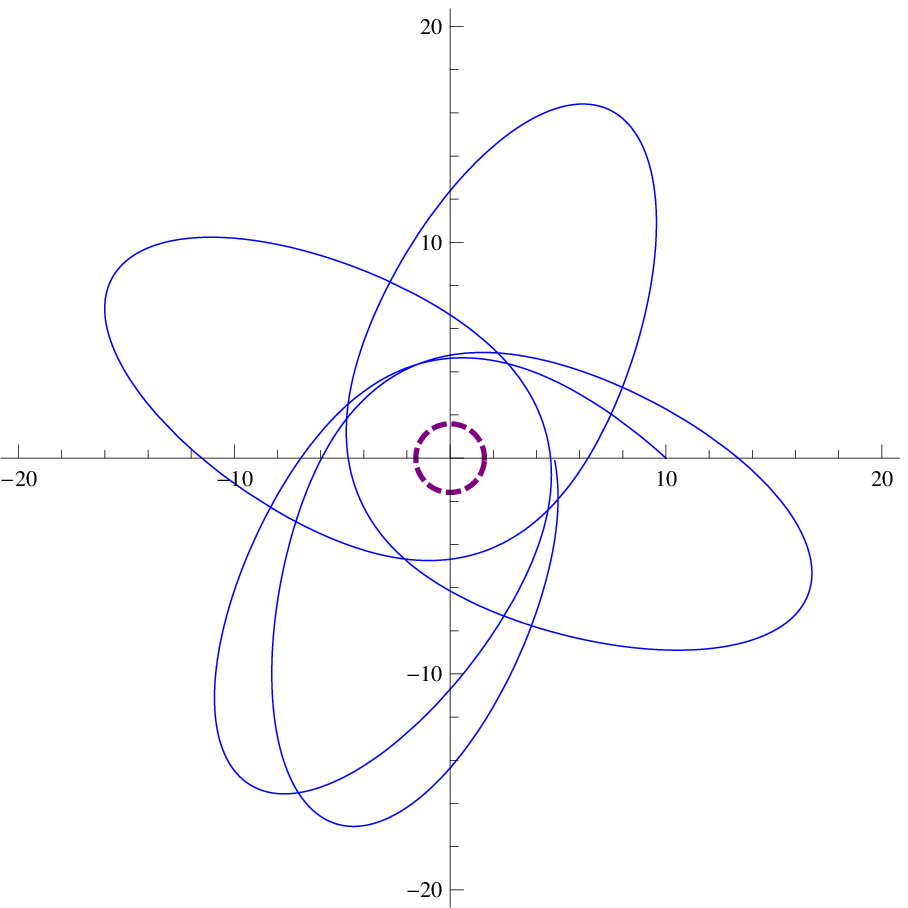}\qquad
\includegraphics[scale=0.55]{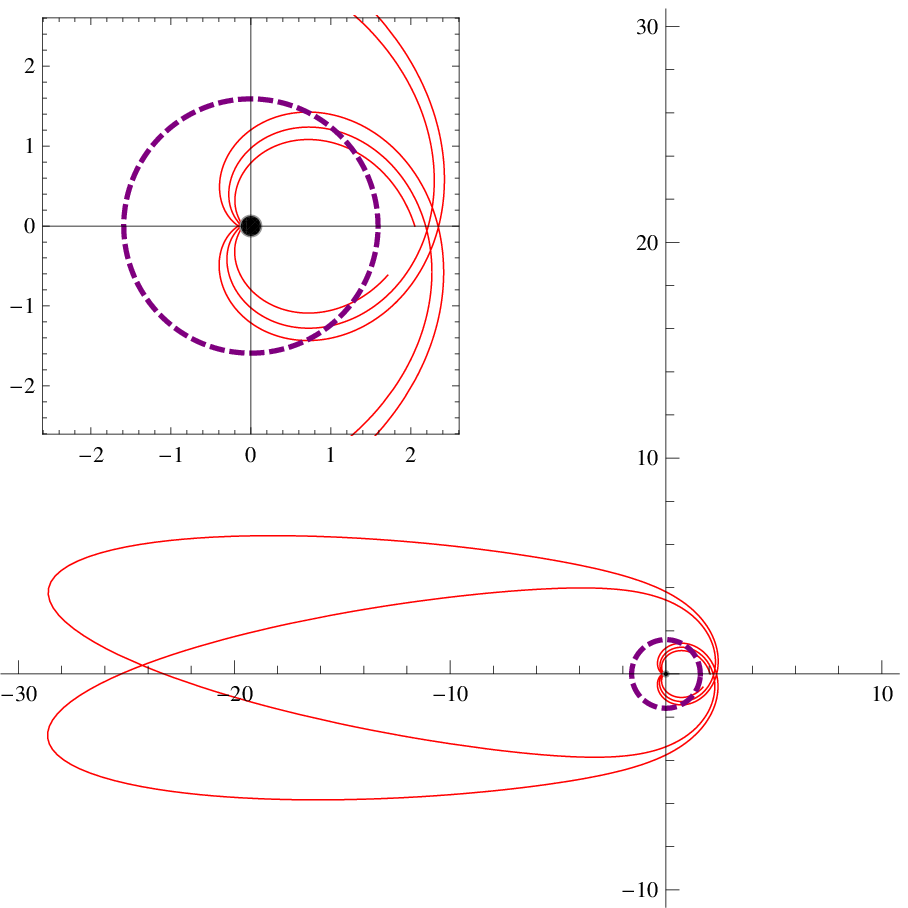}
\caption{Examples of possible timelike trajectories corresponding to a particular profile of $V_{eff}$. Top: the black line corresponds to the effective potential with fixed $G = M=1,Q=0.5,\Lambda=10^{-5},\gamma=0.1,L^2=40$. $E_{tot}$ is shown by two horizontal lines while dashed blue line corresponds to $E^2=3$ and dotted-dashed red line is $E^2=4$. Bottom-left: an orbit and its precession corresponding to $E^2=3$. Bottom-centre: a larger orbit with $E^2=3$. Bottom-right: a mixed bounded orbit with $E^2=4$. The dashed purple circles indicate the outer event horizon of the black hole while the black circle indicates the inner event horizon.}
\label{fig:Parfig}
\end{figure*}

The situation is quite different for timelike trajectory. As shown in the upper panel in Fig. \ref{fig:Parfig}, the corresponding effective potential is distinct from the null case. This can be seen through Eq. (\ref{Veff}). For the timelike case ($e=-1$), there is a scale where $+\gamma r$ dominates, resulting in increasing of effective potential as $r$ increases. It should be remarked that, at the scale where $\Lambda r^2$ dominates, the potential begins to decrease but such a scale is not shown in the potential profile in Fig. \ref{fig:Parfig}.

 
Given that $E_{tot}$ is smaller than the local maximum of $V_{eff}$, as indicated by the dashed blue line in the upper panel of Fig. \ref{fig:Parfig}, there are two possible timelike orbits. 
In similar to the null case, when $r(0)<r_{max}$, massive particle travels in a bounded orbit with precession once it completes its orbit. In the case where $r(0)>r_{max}$, then the massive particle takes a larger orbit around the black hole as shown in the lower centre panel in Fig. \ref{fig:Parfig}. The particle is bounded by gravitational potential outside the black hole. We also observe the precession of its orbit. This bounded orbit outside the black hole's outer horizon occurs because there is a potential well at $r>r_{max}$ which corresponds with the non-vanishing $e$ in Eq. (\ref{Veff}).
  
Lastly, if the corresponding $E_{tot}$ is larger than the local maximum, the resulting trajectory of the particle is somehow a mixture of both the small and the large bounded orbits mentioned earlier. The mixing bounded orbit is illustrated in the lower-right panel of Fig. \ref{fig:Parfig}. In particular, the particle travels inward and outward through the outer horizon when it gets close to the black hole and orbits with larger radial coordinate when it is far from the black hole. At very large $r$, the contribution from the angular momentum $L$ is very small according to Eq. (\ref{Veff}). Thus, the trajectory of a massive particle is essentially a straight line. Note that, it is possible for timelike particle to have a deflected fly-by trajectory when $r$ is large and $L$ is non-negligible.


\section{Conclusions}
\label{conclusions}

In this work, we have investigated the gravitational lensing phenomenon and particle motions around a black hole in the presence of massive graviton. We particularly focus on the dRGT massive gravity. In dRGT massive gravity, the black hole solutions differ from conventional black holes in GR, namely, there are a cosmological constant-like term, a linear $\gamma$ term, and a global monopole term which originate from the massive graviton. We find deflection angle formula of light  analytically, where the formula is divided into two classes: large and small $\gamma$ term. Our formulae suggest that deflection angles in dRGT massive gravity can either be positive, zero and negative. The negative deflection angle is interpreted as gravity works repulsively. Both analytic formulae are confirmed by direct numerical integration. For the small $\gamma$ regime, we have applied our formula to calculate gravitational lensing effect on the SMBH of the Milky Way, Andromeda galaxy, and Virgo A galaxy. The results reveal that the deflection angle with massive graviton is smaller than that with massless graviton case at large distances because of the linear $\gamma$ term. The linear $\gamma$ term yields repulsive behaviour of gravity such that at large distances the deflection angle can be negative. This behaviour is confirmed in both regimes. 

Another aspect of this work is to explore how a particle moves in the dRGT massive gravity. Therefore, we study the null and timelike geodesics around the dRGT black hole. In both cases, geodesic equations are numerically solved and trajectories of particle are displayed. From our investigation, there are three possible trajectories for photon, i.e. bounded orbit, attractive fly-by and repulsive fly-by. For the bounded orbit, photon is trapped by gravitational potential of the black hole. The photon travels through the outer horizon and get bounced back at the inner horizon. Both types of fly-by trajectories occur at large distance from the black hole. We find that if the distance is considerably far from the black hole, dRGT black hole acts as a concave lens where light ray is diverged rather than focused. This agrees with the repulsive behaviour of gravity that we obtain from our deflection angle formulae.


For the case of a massive particle, we investigate bounded trajectories. The bounded orbits can be classified into three categories, i.e. small, large and mixed orbits. For the small bounded orbit, the particle travels too close to the black hole and get trapped by gravitational pull of the black hole. The similar motion is also found for a larger orbit. There also exists a mixing bounded orbit where the particle orbits around the black hole in both small and large orbits.

To extend this work further, one can consider the following effects on dRGT black hole, black hole shadow, Einstein ring and Shapiro time delay. On the other hand, since this work focuses purely on static spherically symmetric black hole solution, it is interesting to investigate the similar local gravity phenomena, i.e. gravitational lensing, in the context of cylindrically symmetric or axially symmetric setup.


\section*{Acknowledgement}

S.P. (first author) is supported by Rachadapisek Sompote Fund for Postdoctoral Fellowship, Chulalongkorn University. L.T. is supported by the National Research Foundation of Korea (NRF) grant funded by the Korea government (MSIP) (Grant No. 2016R1C1B1010107).

\appendix

\section{Parameter Space of Charged dRGT black hole}
\label{chargedBH}

Here in this section, we explore parameter space of a charged dRGT black hole. Despite the solution Eq. (\ref{drgtmetric}) can be expressed in either positive or negative cosmological constant, we shall only consider in the case where $\Lambda$ is positive in this work. With this choice, the spacetime metric Eq. (\ref{drgtmetric}) becomes a modified Reissner-Nordstr\"om de-Sitter spacetime. Generally speaking, the metric function Eq. (\ref{drgtmetric}) has four real roots. However, only a positive root will be considered as a horizon. A quick investigation \cite{Burikham:2017gdm} reveals that there are several possible scenarios e.g. three positive roots $r_-<r_h<r_c$, two positive roots $r_-=r_h<r_c$ or $r_-<r_h=r_c$ (extremal case) and one positive root. Remark that, the Cauchy, event and cosmological horizons are denoted by $r_-,r_h$ and $r_c$, respectively. It should be noted that for the remaining part of this article, we shall set $\zeta=0$ and $G=1$ unless otherwise stated.

Now, we shall explore roots structure of charged dRGT black holes Eq. (\ref{drgtmetric}) in more details. We generalise the analysis done for parameter space of Reissner-Nordstr\"om de-Sitter black hole in \cite{Mokdad:2017udm}. The analysis of black hole's horizon structure (with strictly positive $\Lambda$) can be done by considering the following arguments. First let's define a new polynomial function, its first and second derivatives
\begin{align}
H &= r^2 f = -\frac{\Lambda r^4}{3} + \gamma r^3 + r^2 -2M r + Q^2, \\
H'&= -\frac{4\Lambda r^3}{3} + 3 \gamma r^2 + 2r - 2M, \\
H'' &= -4\Lambda r^2 + 6\gamma r +2,
\end{align}
where ``prime" is derivative with respect to $r$.
Equating the last equation to zero yields
\begin{align}
R_{\pm} &= \frac{3\gamma\pm\sqrt{9\gamma^2+8\Lambda}}{4\Lambda}.
\end{align}
These are the extremum point of $H'$. We shall consider only $R_+$ since it is always positive. Since $H''(0)=2$ and $H''(R_+)=0$, hence $H'$ is increasing in the interval $r\in [0,R_+)$ and has local maximum at $R_+$. Then $H'$ is decreasing in the interval $r\in (R_+,\infty)$. Assuming $H'(R_+)<0$, then $H'$ is negative for $r>0$ since $H'(0)<0$. Therefore $H$ is decreasing on $r\in[0,\infty)$. In this case, $H$ has only one positive root since $H(0)=Q^2 > 0$. \\

\begin{figure}[b]
\centering
\includegraphics[width=0.45\textwidth]{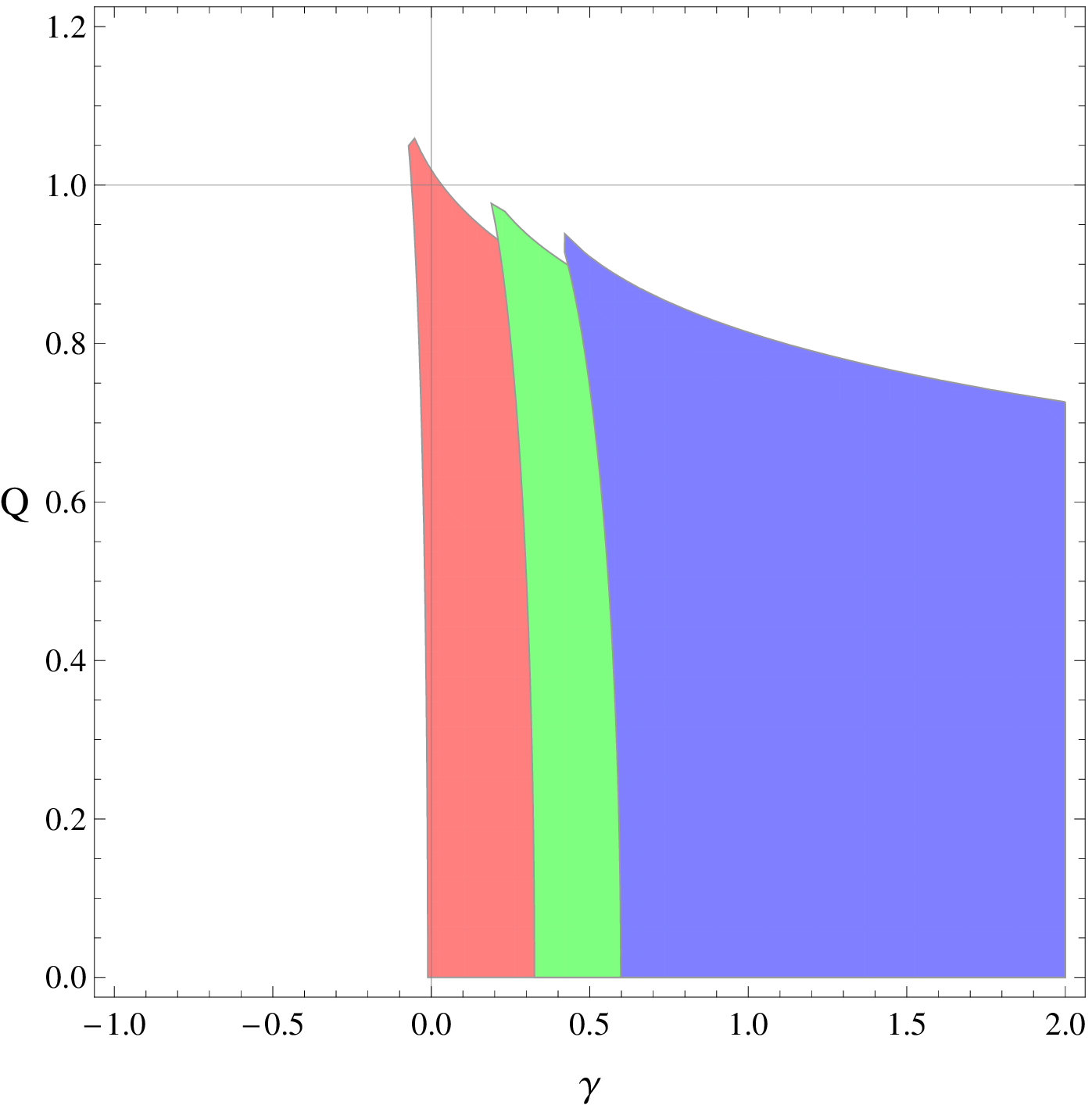}\hspace{5mm}
\includegraphics[width=0.45\textwidth]{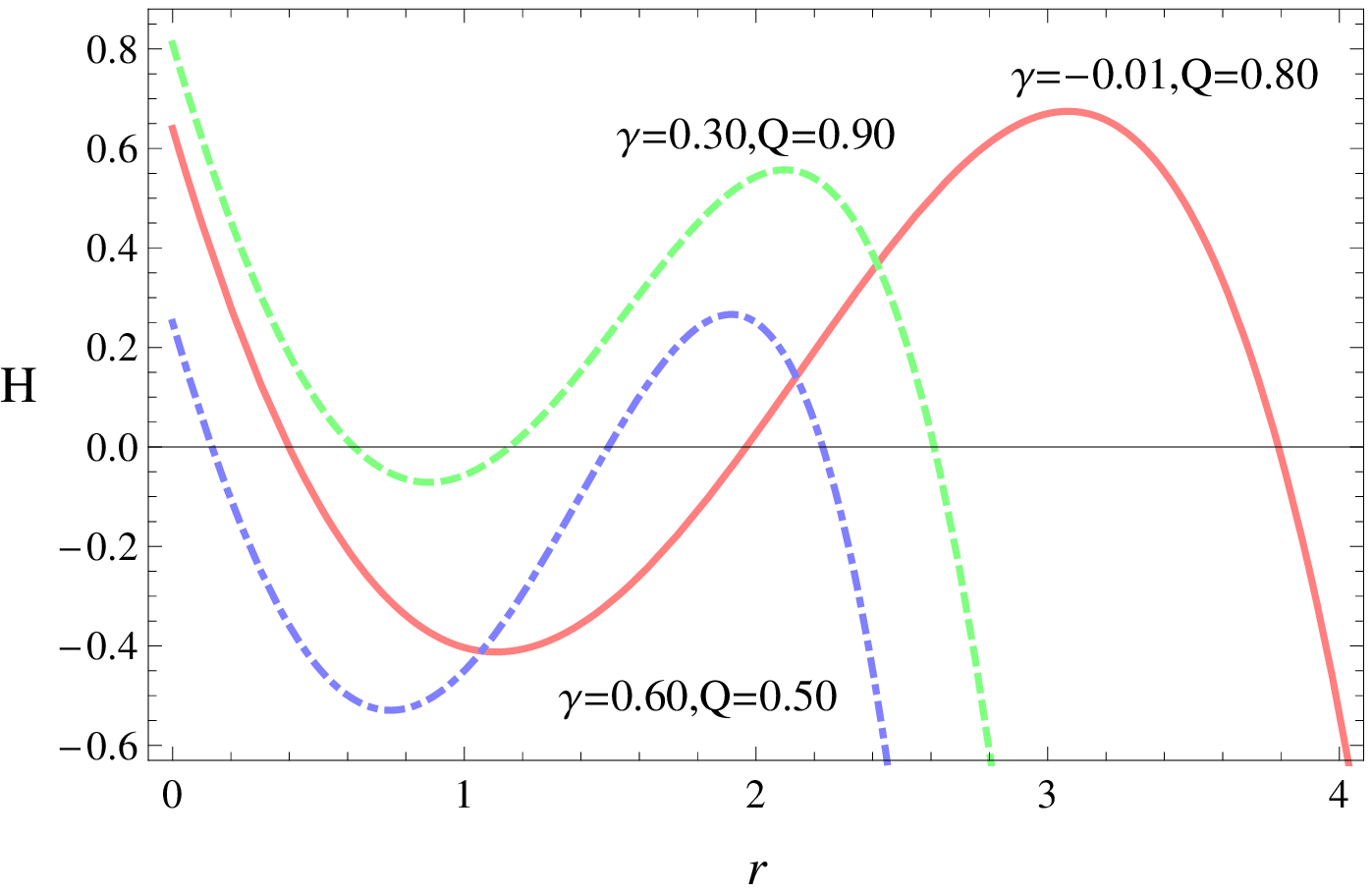}
\caption{Top: Parameter space of black holes with three horizons. Bottom: Example plots of $H$ with three positive roots. Colour code: (Large, Solid, Red) $\Lambda=0.1$, (Medium, Dashed, Green) $\Lambda=0.5$ and (Small, Dotted-dashed, Blue) $\Lambda=0.9$ for $M=1$. See online version for colour.}
\label{fig:appfirst}
\end{figure}

Since $\Lambda>0$, we expect Eq. (\ref{drgtmetric}) to have two (extremal case) or three positive roots. These can be obtained by considering the condition $H'(R_+)>0$, which is equivalent to
\begin{align}
\frac{R_+}{6}\left(6+9\gamma R_+ - 4\Lambda R_+^2\right) > M. \label{Hcond}
\end{align}
We know that $\lim\limits_{r\to\pm\infty} H'(r) = \mp \infty$, $H'(0)<0$ and $H'$ has a positive local maximum at $R_+$. 
Therefore $H'$ must has two positive roots and one negative root. Now we define two positive roots of $H'$ as $h_1,h_2$ such that $0<h_1<R_+<h_2$. Hence over the interval $r\in[0,\infty)$, we now have
\begin{align}
H'(h_1)&=0,~H''(h_1)>0,\Rightarrow H(h_1)~\text{is local minimum}, \\
H'(h_2)&=0,~H''(h_2)<0,\Rightarrow H(h_2)~\text{is local maximum}.
\end{align}
Three positive roots of Eq. (\ref{drgtmetric}) shall be denoted by $r_1 < r_2 < r_3$. Finally since $H(0)=Q^2 > 0$, the root structure of Eq. (\ref{drgtmetric}) can be classified according to $H(h_1)$ and $H(h_2)$ as follows
\begin{itemize}
\item For $H(h_1)>0$, there is only one positive root at $r_1>h_2$.
\item For $H(h_1)=0$, there are two positive roots at $r_1=r_2=h_1$ and $r_3 > h_2$. This is an extremal case where black hole's Cauchy horizon and event horizon exist at the same location.
\item For $H(h_1)<0$, there are three possible scenarios.
\begin{itemize}
\item If $H(h_2)<0$, $H$ has only one root at $r_1<h_1<h_2$.
\item If $H(h_2)=0$, $H$ has two roots at $r_1<h_1$ and $r_2=r_3=h_2$. This is another extremal case where black hole's event horizon and cosmic horizon coincide.
\item If $H(h_2)>0$, $H$ has three positive roots at $r_1<h_1<r_2<h_2<r_3$. This is the case where Eq. (\ref{drgtmetric}) possesses Cauchy horizon, event horizon and cosmological horizon.
\end{itemize}
\end{itemize}
We therefore conclude that the conditions necessary for three horizons case are 
\begin{align}
H'(R_+)>0,~~~~H(h_1)<0,~~~~H(h_2)>0. \label{Hcond2}
\end{align}

In Fig.~\ref{fig:appfirst} (Top), an example of parameter space for black holes with three horizons are displayed. These parameter spaces are plotted such that conditions in Eq. (\ref{Hcond2}) are simultaneously satisfied. Each plots, the black hole mass $M$ is set to be unity and cosmological constant is fixed to be $0.1,0.5$ and $0.9$. The allowed region becomes smaller as $\Lambda$ increases. For each fixed $\Lambda$, the lower bound of $\gamma$ is governed by the condition $H'(R_+)>0$ (or equivalently Eq. (\ref{Hcond})). From these plots, three horizons black holes with negative $\gamma$ only exist in small $\Lambda$ regime. It should be emphasised that the allowed region can be extended further to a larger $\gamma$. In addition, example plots of $H$ which has three positive real roots are shown in Fig.~\ref{fig:appfirst} (bottom). These three roots can be associated with black hole's inner, outer and cosmic horizons. We use colour to distinguish the difference in the value of $\Lambda$. 

\begin{figure}[t]
\centering
\includegraphics[width=0.45\textwidth]{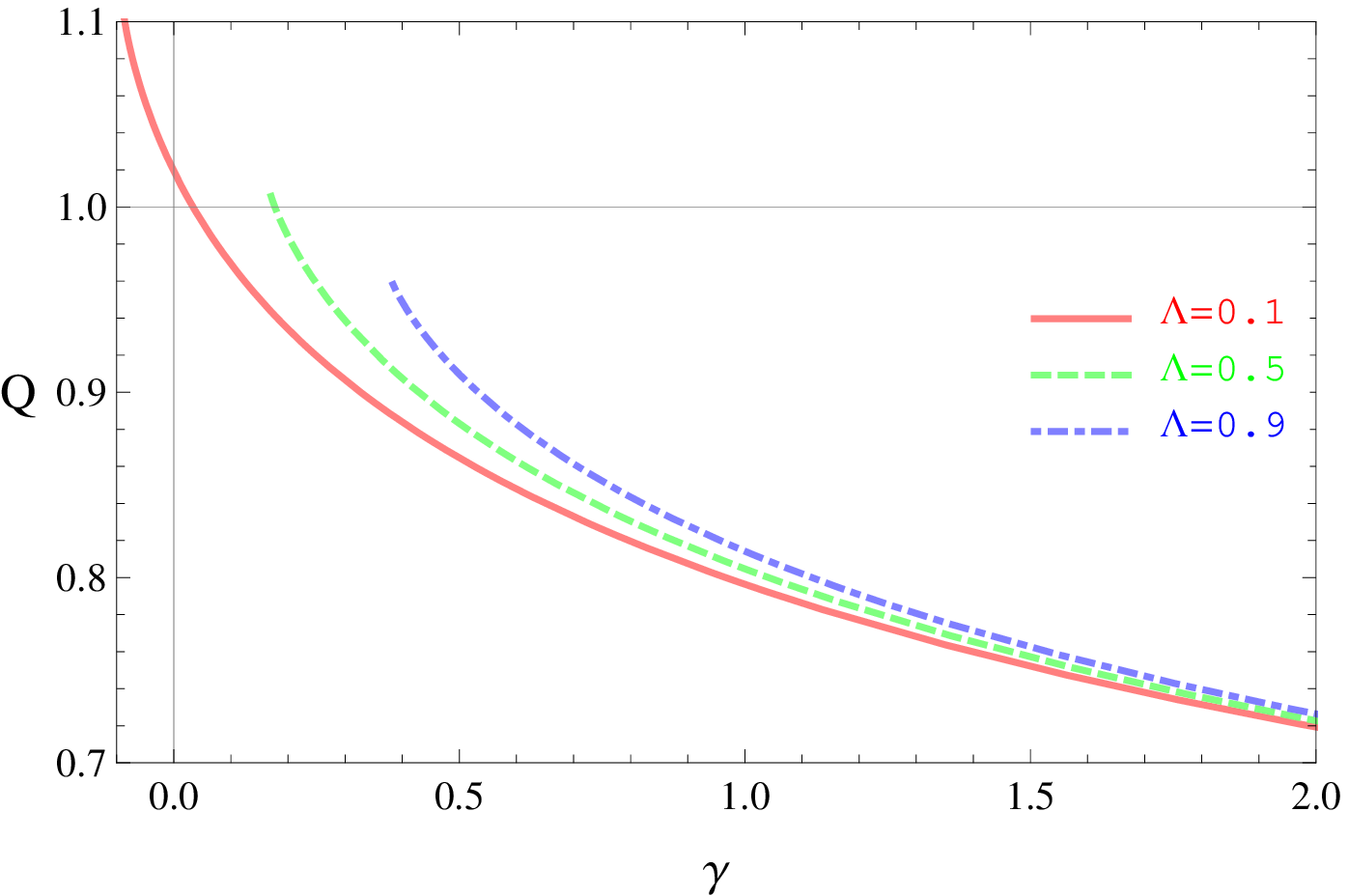}\hspace{5mm}
\includegraphics[width=0.45\textwidth]{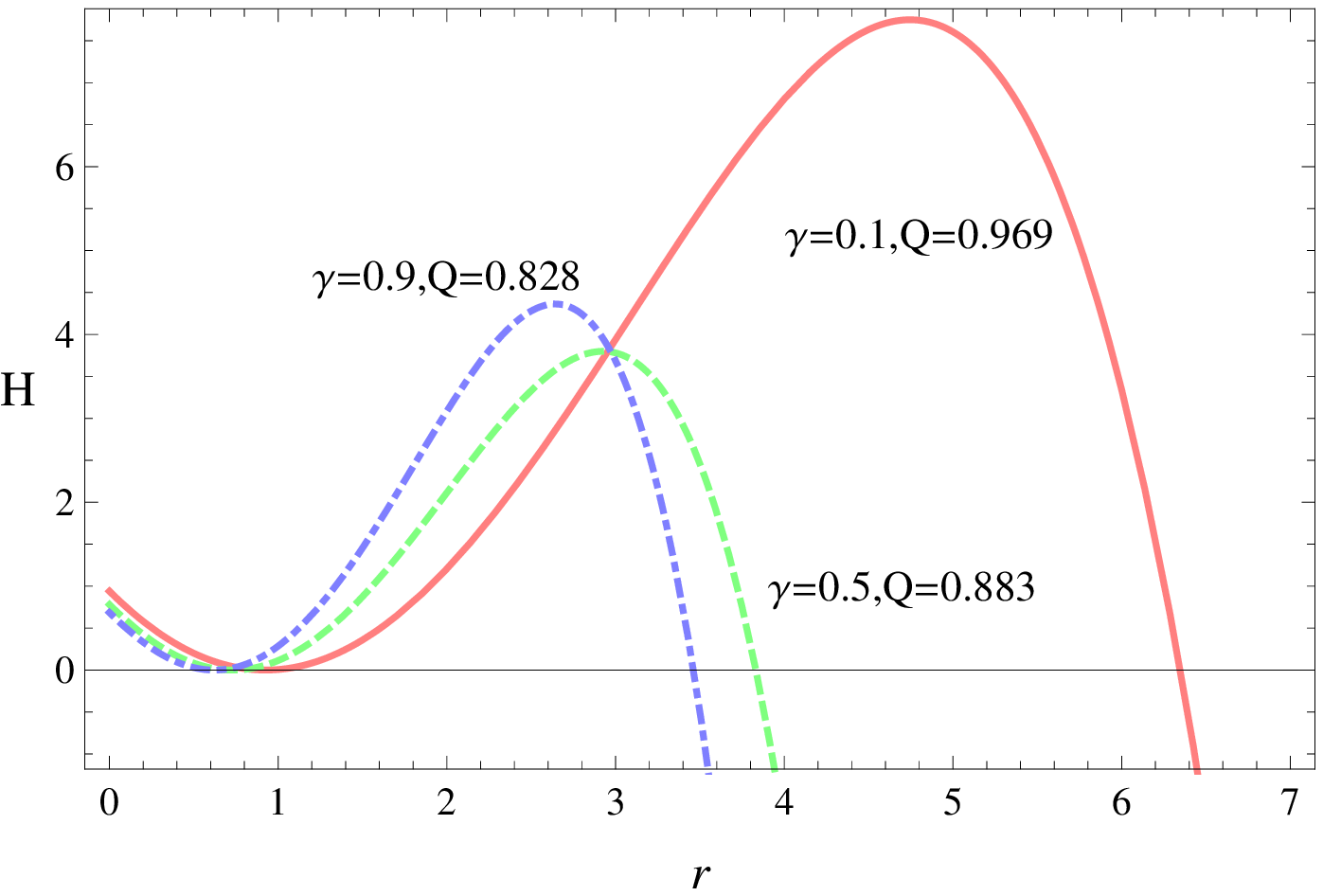}
\caption{Top: Phase space of extremal black holes with $r_-=r_h$. Bottom: Example plots of $H$ with two positive roots. Colour code: (Large, Solid, Red) $\Lambda=0.1$, (Medium, Dashed, Green) $\Lambda=0.5$ and (Small, Dotted-dashed, Blue) $\Lambda=0.9$ for $M=1$. See online version for colour.}
\label{fig:appsecond}
\end{figure}

\begin{figure}[b]
\centering
\includegraphics[width=0.45\textwidth]{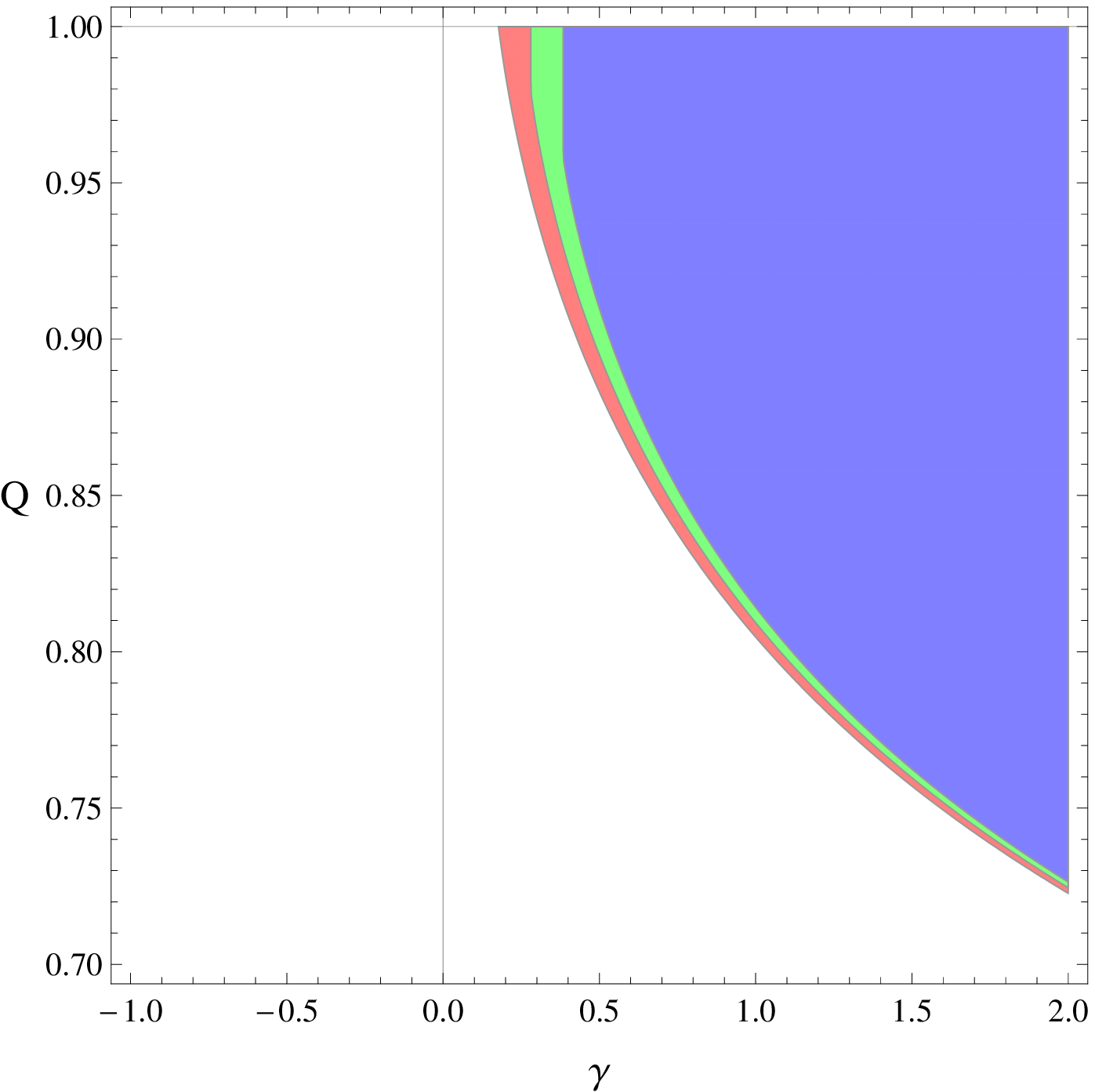}\hspace{5mm}
\includegraphics[width=0.45\textwidth]{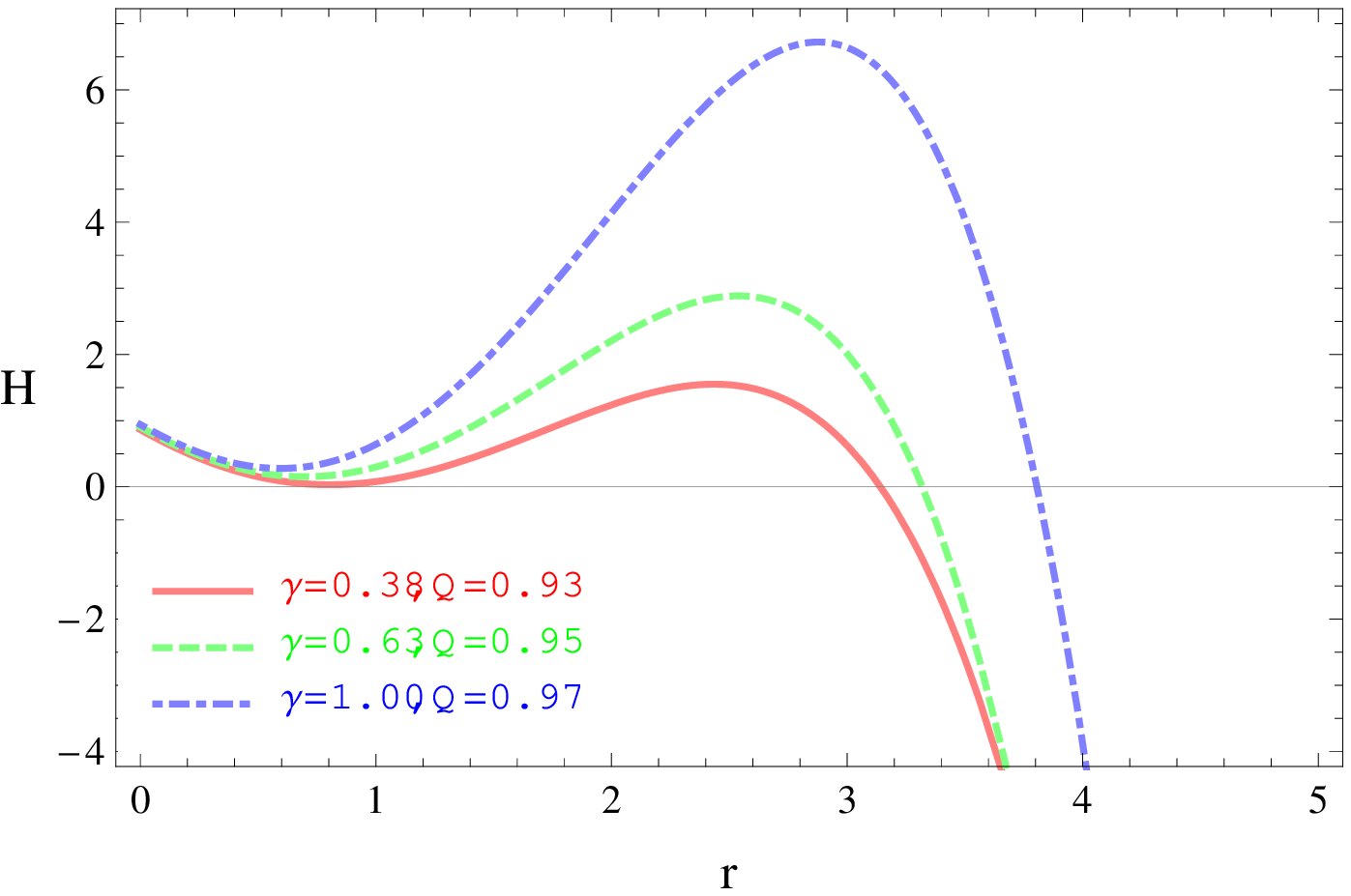}
\caption{Top: Phase space of black hole with one horizon. Bottom: Example plots of $H$ with one positive root. Colour code: (Large, Solid, Red) $\Lambda=0.5$, (Medium, Dashed, Green) $\Lambda=0.7$ and (Small, Dotted-dashed, Blue) $\Lambda=0.9$ for $M=1$. See online version for colour.}
\label{fig:appthird}
\end{figure}

Another interesting aspect of this analysis is an extremal case where the Cauchy horizon coincides with the event horizon. The necessary conditions for this scenario are the following: $H'(R_+)>0$ and $H(h_1)=0$. For fixed $\Lambda$, the first condition gives the constraint on $\gamma$ (recall that this is equivalent with Eq. (\ref{Hcond})). Moreover, the second condition equates $\gamma$ and $Q$ when $\Lambda$ is fixed. For this reason, the phase space of extremal black hole is a line rather than a region and this is illustrated in Fig.~\ref{fig:appsecond}. First of all, negative-$\gamma$-extremal solution is possible for small cosmological constant. As $\Lambda$ increases, the lower bound of $\gamma$ also increases. We also observe the general trend that as $\gamma$ increases the black hole charge must decrease. From numerical investigation, we find that as $\gamma$ increases the lowest possible value of $Q$ approaches some constant. Moreover, three example plots of $H$ which has two positive real roots are displayed in Fig.~\ref{fig:appsecond} (bottom). Let's emphasise that, there is another possibility of extremal case where $r_h=r_c$. However, we choose to omit this scenario since it lies beyond the scope of this paper.

For the sake of completeness, we display an example plot of phase space for a black hole with one horizon. Regarding to our analysis, there are three possible conditions for one positive root. We pick one condition for demonstration purpose that is $H'(R_+)>0$ and $H(h_1)>0$. In Fig. \ref{fig:appthird}, we show the phase space of such black holes (top) and behaviours of $H$ (bottom). We observe that the area of the phase space decreases as cosmological constant $\Lambda$ increases. For a given $\gamma$, there exists some lower bound of $Q$ for which the solution exists. In addition when $\gamma$ increases, the minimum value of $Q$ becomes lower. One final remark is, the other two conditions give a large region of parameter space where black hole has one horizon with $\gamma<0$.


\end{document}